\documentclass[aps,prd,reprint,twocolumn,superscriptaddress,longbibliography,nofootinbib,floatfix,showpacs]{revtex4-1}
\usepackage{epsfig}
\usepackage{amsmath,amssymb,amsfonts}
\usepackage[breaklinks]{hyperref}
\usepackage{mathrsfs}
\usepackage{bbm}
\usepackage{slashed}
\usepackage{graphicx}
\usepackage{verbatim}

\graphicspath{{./fig/}}

\usepackage[usenames]{color}

\definecolor{darkgreen}{rgb}{0.2,0.6,0}

\newcommand{\Eqref}[1]{Eq.~\eqref{#1}}

\newcommand{\be}{\begin{equation}}
\newcommand{\ee}{\end{equation}}
\newcommand{\bw}{\begin{widetext}}
\newcommand{\ew}{\end{widetext}}
\newcommand{\bi}{\begin{itemize}}
\newcommand{\ei}{\end{itemize}}
\newcommand{\bea}{\begin{eqnarray}}
\newcommand{\eea}{\end{eqnarray}}
\newcommand{\ud}{\mathrm{d}}

\newcommand{\ImG}{\text{Im }\Gamma}
\newcommand{\LCm}{{\scriptscriptstyle -}} 
\newcommand{\LCp}{{\scriptscriptstyle +}}
\newcommand{\LCpm}{{\scriptscriptstyle \pm}}

\newcommand{\LCperp}{{\scriptscriptstyle \perp}}

\usepackage[T1]{fontenc} \usepackage[utf8]{inputenc}

\begin{document}

\title{Critical Schwinger pair production II - universality in the deeply critical regime}

\author{Holger Gies}
\email[]{holger.gies@uni-jena.de}
\affiliation{Theoretisch-Physikalisches Institut, Abbe Center of Photonics, Friedrich-Schiller-Universit\"at Jena, 
Max-Wien-Platz 1, D-07743 Jena, Germany}   
\affiliation{Helmholtz-Institut Jena, Fr\"obelstieg 3, D-07743 Jena, Germany}

\author{Greger Torgrimsson}
\email[]{greger.torgrimsson@uni-due.de}
\affiliation{Fakult\"at f\"ur Physik, 
Universit\"at Duisburg-Essen, Lotharstra{\ss}e 1, Duisburg 47048, Germany}

\begin{abstract}
We study electron-positron pair production by spatially inhomogeneous
electric fields. Depending on the localization of the field, a
critical point (critical surface) exists in the space of field
configurations where the pair production probability vanishes. Near
criticality, pair production exhibits universal properties similar to
those of continuous phase transitions. We extend results previously
obtained in the semi-classical (weak-field) critical regime to the
deeply critical regime for arbitrary peak field strength. In this
regime, we find an enhanced universality, featuring a unique critical
exponent $\beta=3$ for all sufficiently localized fields. For a large
class of field profiles, we also compute the non-universal amplitudes.
\end{abstract}

\maketitle
\section{Introduction}

Universality is a paradigm that often arises from the dominance of
long-range fluctuations near critical points, washing out the
effect of microscopic details on the long-range observables. This form
of universality can be cast into scaling laws of observables which are
characterized by universal critical exponents that depend only on a
few gross features of the system such as dimensionality, symmetries
and the number and nature of the long-range degrees of freedom.
Standard examples are provided by critical phenomena in spin systems
or liquid-gas transitions
\cite{Kadanoff:1971pc,Wilson:1973jj,ZinnJustin:2002ru}.

Beyond fluctuation dominated systems, universality has also become a
useful concept in classical (deterministic) systems such as turbulence
\cite{Kolmogorov9}, or even general relativity
\cite{Choptuik:1992jv,Gundlach:2007gc}, where the resulting scaling
laws reflect self-similarity of the field configurations induced by
the nonlinearities of the underlying theory over a wide range of
scales.

In a recent letter~\cite{Gies:2015hia}, we have found aspects of
universality also in Schwinger pair
production~\cite{Sauter:1931zz,Heisenberg:1935qt,Schwinger:1951nm},
where an analogue of a critical point exists in the form of field
configurations that provide the minimum of electrostatic energy to
produce a (real) pair from vacuum, see
e.g.~\cite{Wang:1988ct,Gies:2005bz,Hebenstreit:2011wk,Polyakov-conjecture}. On the one hand, this form of
universality appears to fit into the framework of fluctuation-driven
criticality, as the onset of pair production arises from long-range
electron-positron quantum fluctuations that acquire sufficient energy
from the external field to become real. On the other hand, the
relevant physics of this process can be extracted from the
Klein-Gordon or Dirac equation in an external field, which may be
viewed as a classic deterministic and even linear wave equation.  

This makes universality in Schwinger pair production a rather special
example. Nevertheless, the origin of this universality has a clear
physical picture: the relevant long-range fluctuations average over
the local details of the pair-producing field profile, giving rise to
a scaling law that depends only on the large-scale properties of the
field. We emphasize that this universality holds only near
criticality. By contrast, the microscopic details of the field can
play an important role in other parameter regions, such as in the
dynamically assisted pair production
regime\cite{Schutzhold:2008pz,Schneider:2014mla,Linder:2015vta,%
  Orthaber:2011cm,Hebenstreit:2009km,DiPiazza:2009py,Dumlu:2010vv,Jansen:2013dea,Akal:2014eua,Otto:2015gla,Panferov:2015yda}. The
diversity of this phenomenon of pair production and its interpretation
as a decay of the vacuum have lead to a search and study of analogue
systems in e.g. atomic ionization \cite{Popov:2004}, graphene
\cite{Allor:2007ei,Fillion-Gourdeau:2015dga}, and
semiconductors
\cite{Linder:2015fba,Oka:2005} and with ultracold atoms in optical
lattices
\cite{Szpak:2011rs,Szpak:2011jj,Queisser:2012aa,Kasper:2015cca}.

In our previous work~\cite{Gies:2015hia}, we interpreted the pair
production probability $\ImG$ as an order parameter and determined the
scaling of this order parameter with the distance from the critical
point. Using semiclassical worldline instanton
methods~\cite{Affleck:1981bma,Dunne:2005sx,Dunne:2006st,Dunne:2006ur},
we found that the \textit{semiclassical critical regime} entails a
family of critical exponents $\beta$ which is directly related to the
power by which the electric field vanishes, i.e. the power $d$ for
asymptotically vanishing fields $E\sim x^{-d}$, or $n$ for fields with
compact support $E\sim (x-x_0)^n$. For each $d$ and $n$ there is one
universality class. Though the worldline instanton approach
facilitates a direct understanding of criticality and universality,
and provides quantitative information about the semiclassical region near the critical point, the critical
point itself actually lies outside the semiclassical regime of
validity. This makes the large degree of universality that we found
even more remarkable. It is natural to expect that universality will
be enhanced in the immediate vicinity of the critical point. In
this paper, we verify this expectation in the affirmative.

This paper is organized as follows. In Sect.~\ref{Enhanced
  universality} we give a general introduction as well as a brief
summary of the main results. In Sect.~\ref{Derivation of the universal
  critical exponent} we derive the universal critical scaling of the
probability for fields vanishing asymptotically faster than
$|x|^{-3}$, and in Sect.~\ref{D for strong fields} and Sect.~\ref{D
  for weak fields} we derive also the non-universal coefficient under
the additional assumption of either strong or weak field strengths. In
Sect.~\ref{Fields decaying as x3} and Sect.~\ref{Fields decaying as
  x2} we study fields that decay as $|x|^{-3}$ and $|x|^{-2}$,
respectively. In Sect.~\ref{Spinor QED}, we briefly consider spinor
QED. We conclude in Sect.~\ref{Conclusion}.

\section{Universality in Schwinger pair production}\label{Enhanced universality}

Schwinger pair production denotes the instability of the asymptotic
``in'' vacuum towards the creation of pairs in the presence of an external
electric or electromagnetic field. The decay probability $P$ of the
vacuum is related to the imaginary part of the QED effective action
$\Gamma$, 
\begin{equation}
P=1-\exp(-2\,\ImG[E]) \;.
\end{equation}
To lowest order, $\ImG$ hence is a measure for the pair production
rate \cite{Nikishov:1970br,Nikishov:2001ps,Cohen:2008wz}. From the viewpoint of
critical phenomena, we consider $\ImG$ as an order parameter for pair
production. In the space of all conceivable electromagnetic field
configurations, $\ImG$ can only be nonzero, if the external background
can transfer sufficient energy to the electron-positron fluctuations
to form a real pair. In the infinite dimensional space of field
strength tensor functions, the regions where $\ImG\neq0$ are therefore
separated from those where $\ImG=0$ by a critical hypersurface. 

In the present work, we confine ourselves to a large class of field
configurations within which we can approach the critical surface from
the unstable-vacuum side ($\ImG\neq0$) by tuning one parameter. For
this, we consider unidirectional spatially inhomogeneous electric
background fields with one nonzero vector component $E(x)$, which
varies along the direction $x$ of the field. For convenience, we use
units with $\hbar=c=1$ and absorb a factor of the electron charge into
the background field, $eE\to E$. We concentrate on pair production to
leading-order, ignoring radiative corrections of the photon field
which would involve higher orders in the fine-structure constant
$\alpha=e^2/(4\pi)$, see, e.g.
\cite{Ritus:1975cf,Dittrich:1985yb,Lebedev:1985bj,Fliegner:1997ra,%
  Kors:1998ew,Dunne:1999vd,Dunne:2002qg,Dunne:2004xk,Huet:2010nt,Huet:2011kd,Schubert-conjecture}.
We also use units in which the rest mass of the electron is set to
$m=1$, implying that all dimensionful quantities are expressed in
units of the electron mass.

The fields of interest can be parametrized by $E=A'$ with potential 
\begin{equation}
A(x)=\frac{1}{\gamma}(1+f(kx)), \quad \gamma= \frac{k}{E_0} \;,\label{eq:A}
\end{equation}
where $E_0$ is a characteristic field strength scale and $k^{-1}$ a
characteristic length scale of the inhomogeneous field. Their precise
choice is not relevant. In fact, the field profile may support various
of these scales, such that the function $f(kx)$ in addition depends on
dimensionless ratios of further scales. Of particular relevance is the
adiabaticity parameter $\gamma=k/E_0$, as we limit ourselves to fields
with 
\begin{equation}
\label{1f1} 
f(-\infty)<f(kx)<f(\infty) 
\end{equation}
and normalize the profile function $f$ such that
$f(\pm\infty)=\pm1$. As a consequence, $A(-\infty)=0$ and
$A(+\infty)=2/\gamma$. This class of fields goes beyond those
considered frequently in the literature, in particular there is no
restriction concerning monotonicity and (anti-)symmetry of $f$.

A semi-classical viewpoint suggests that pair production requires
the electric field be sufficiently strong or extended to provide an
electrostatic energy greater than the energy of a real pair at
rest. In the full quantum theory, this threshold may receive quantum
radiative corrections from final-state interactions, which come, however, with higher powers of $\alpha$.
Recalling that $m=1$, the energy constraint reads
\begin{equation}
\int\limits_{-\infty}^\infty\!\ud x\; E>2 \implies
\gamma<1 \;.\label{eq:E}
\end{equation}
In our previous paper \cite{Gies:2015hia}, we have studied criticality in the semiclassical
regime, 
\begin{equation}
E_0^2\ll 1-\gamma^2\ll1 \;. \label{eq:semicrit}
\end{equation}
We have considered fields that decay asymptotically with a power law
$E\to E_0c(kx)^{-d}$ or vanish at a finite point $x_0$ as $E\to E_0c(k[x-x_0])^n$. In the
semiclassical regime, we have \cite{Dunne:2006st,Gies:2015hia}
\begin{equation}
\text{Im }\Gamma\sim\frac{\exp\big[-\frac{\pi}{E}g(\gamma^2)\big]}{(\gamma^2 g)'\sqrt{(\gamma^2 g)''}} \qquad (...)':=\frac{\ud}{\ud\gamma^2}(...)
\end{equation}
and, for $n>1$ and $d>3$, the critical limit is obtained from
\begin{equation}\label{semiclassical-g}
g=\frac{2}{\pi}\!\int\limits_{-\infty}^\infty\!\ud u\sqrt{1-f^2(u)}+C(1-\gamma^2)^\rho+... \;,
\end{equation}
where $C(s,c)$ and $\rho=\frac{1}{2}\frac{s+3}{s+1}$, with $s=n,-d$,
only depend on the asymptotic behavior of the field. The scaling for
$n\leq1$ and $d\leq3$ can be obtained from second term in
\Eqref{semiclassical-g} and its first and second derivative,
see~\cite{Gies:2015hia}. 
Differentiating the second term in \Eqref{semiclassical-g} gives terms that cause the prefactor of $\text{Im }\Gamma$ to vanish as $\gamma\to1$. Although $\rho>0$ for $d>3$, the second term can also be quantitatively relevant in the exponent for $1-\gamma^2>E_0^2$. For $\rho<0$ we find essential scaling. Thus, the scaling is determined by the second term in \Eqref{semiclassical-g}, and hence only depends on the asymptotic behavior of the field. 
So, the power
$d$ or $n$ groups fields into different universality classes in the
semi-classical regime.

The semi-classical critical regime defined in \Eqref{eq:semicrit} is
likely to be most relevant to upcoming experiments, as the field
strength or intensity rather than length scales represent the most
challenging issue, e.g. for high-power lasers. Still, limiting the
criticality study to this regime is conceptually not satisfactory, as
the criticality limit, i.e., the approach of the critical surface is
defined by taking $\gamma\to 1$, say for constant $E_0$. It is the aim
of the present work to study scaling in the \textit{deeply critical
  regime}, defined by the regime where $1-\gamma^2$ is smaller than
any other scale, 
\begin{equation}
 1-\gamma^2\ll \{E^2,1/E^2,1\} \;.\label{eq:deepcrit}
\end{equation}
Below, we show that the scaling in this regime depends even less on the
field. In fact, for all $n$ and $d>3$ we find power-law scaling with
the same universal critical exponent, 
\begin{equation}
\text{Im }\Gamma\propto(1-\gamma^2)^3 \;. \label{eq:universalcrit}
\end{equation}
If we further assume that the field is weak, $E^2\ll1$, we find a general expression for the non-universal coefficient 
\begin{equation}
\label{intro Gamma}
\text{Im }\Gamma=P(n,d)(1-\gamma^2)^3e^{-2S} \;,
\end{equation}
where the prefactor $P$ is universal in the sense that it depends only on the asymptotic behavior of the electric field (see below), and the microscopic details of the field are included in the ``tunneling exponent''
\begin{equation}\label{Sint}
S=\int\limits_{-\infty}^\infty\!\ud u\;\frac{\sqrt{1-f^2}}{k} \;,\quad u=kx \;,
\end{equation}
which we recognize from the semiclassical result~\Eqref{semiclassical-g}.
The result \eqref{intro Gamma} holds for different combinations of $d$
and $n$, e.g. fields decaying with $d_\LCp$ for $x\to\infty$ and
$d_\LCm$ for $x\to-\infty$, or decaying with $d$ for $x\to\infty$ and
with $n$ at a finite point. By taking the limit $d\to\infty$ or
$n\to\infty$ we recover the scaling for an exponentially decaying
field, which agrees in particular with the exact result for the Sauter
field \cite{Nikishov:1970br}. We will see below, that the scaling
differs from $(1-\gamma^2)^3$ for weak fields vanishing slower than
$|x|^{-3}$.

Importantly, the critical scaling $(1-\gamma^2)^3$ also holds without
assuming $E\ll1$. Many aspects of criticality of Schwinger
  pair-production are independent of the spin of the created
  particles; hence, it suffices to perform the study for the simpler
  case of scalar QED for most aspects. An interesting difference
  between scalar and spinor QED occurs though for strong fields
  $E^2\gg1$. For scalar QED,
we find for $E^2\gg1$ that the final result
for the order parameter is given by (recall $f^2\leq1$) 
\begin{equation}
\label{final-strong} 
\text{Im }\Gamma_\text{scal}=\frac{L^2T}{48\pi}E_0^2(1-\gamma^2)^3\Bigg(\int\limits_{-\infty}^\infty\!
\ud u\; 1-f^2(u)\Bigg )^{-2} \;.  
\end{equation}
In fact,
\eqref{final-strong} holds also for fields which vanish slower than
$|x|^{-3}$ as long as the integral in \eqref{final-strong}
converges. Strong fields vanishing as $|x|^{-2}$ still have to be
treated separately.

Spinor QED also exhibits the same scaling with $(1-\gamma^2)^3$. Moreover, for strong fields, we find a
remarkably universal expression
\begin{equation}\label{GammaSpinStrong} 
\text{Im }\Gamma_\text{spin}=\frac{L^2T}{96\pi}(1-\gamma^2)^3 \;.
\end{equation}
A comparison of~\Eqref{GammaSpinStrong} and~\Eqref{final-strong}
reveals two important differences: First, in contrast to the
weak-field regime where scalar and spinor QED predict essentially the
same $\text{Im }\Gamma$, here we find $\text{Im
}\Gamma_\text{spin}\ll\text{Im }\Gamma_\text{scal}$ due the factor of
$E_0^2$ in~\Eqref{final-strong}. Second, spinor QED leads to a higher
degree of universality; in fact, the whole
expression~\Eqref{GammaSpinStrong} is universal and does neither
depend on the details of the profile nor on the asymptotic behavior of
the field.

Note that there is no exponential suppression factor
in~\Eqref{final-strong} or~\Eqref{GammaSpinStrong}, as would be
typical for pair production in weak fields (e.g. as in \eqref{intro
  Gamma}). The large-field regime therefore appears most promising for
a future experimental verification of this universal critical
behavior. In scalar QED, the factor of $E_0^2\gg1$
in~\Eqref{final-strong} can further compensate for the critical factor
$(1-\gamma^2)^3\ll1$, which could be important for analog systems.

\section{Derivation of the universal critical exponent}\label{Derivation of the universal critical exponent}

In the present work, we use the classical field equation for an
analysis of the pair production probability,
see~\cite{Nikishov:1970br,Nikishov:2001ps,Holstein:1999ta,Gavrilov:2015yha}
for more details on this formalism.  As the critical scaling is
independent of spin, we will focus on the scalar case and solve
directly the Klein-Gordon equation. Throughout we consider
$1-\gamma^2$ to be the smallest parameter in the problem and we work
to the lowest nontrivial order. As the fields only depend on $x$, the
Klein-Gordon equation can be written
\begin{equation}
\label{Klein}
\Big(\partial_x^2+[p_0-A(x)]^2-m_\LCperp^2\Big)\varphi=0 \;, 
\end{equation}
where $m_\LCperp^2=1+p_\perp^2$ and $p_\perp$ is the momentum
spatially transverse to the field direction. The momentum longitudinal
to the field for $x\to-\infty$ is
\begin{equation}
\label{longp} p^2:=p_0^2-m_\LCperp^2 \;,
\end{equation}
and becomes for $x\to\infty$ 
\begin{equation}\label{longq}
  q^2:=(p_0-2/\gamma)^2-m_\LCperp^2 \;.  
\end{equation} 
Following \cite{Nikishov:1970br,Nikishov:2001ps,Holstein:1999ta,Dumlu:2013pma}, we
are looking for the solution of \eqref{Klein} that behaves
asymptotically as 
\begin{equation}\label{asymptotic}
  Je^{ipx}+Re^{-ipx}\underset{x\to-\infty}{\leftarrow}\varphi(x)\underset{x\to\infty}{\rightarrow}
  e^{iqx} \;.  
\end{equation} 
The imaginary part of the effective action is obtained
\cite{Nikishov:1970br,Holstein:1999ta} by integrating the tunneling
factor 
\begin{equation}\label{scalar T}
  \mathcal{T}=\frac{q}{p}\frac{1}{|J|^2} 
\end{equation} 
over momentum, 
\begin{equation}\label{Gamma p-int} \text{Im
  }\Gamma=\frac{L^2T}{2}\int\frac{\ud^2
    p_\LCperp}{(2\pi)^2}\int\frac{\ud p_0}{2\pi} \; \mathcal{T}
  \;.  
\end{equation} 
We are in the so-called Klein region, where the energy is in the
classical tunnel regime, and the integration limits are obtained from
(c.f. \eqref{longp} and \eqref{longq})
\begin{equation}
  m_\LCperp<p_0<\frac{2}{\gamma}-m_\LCperp \;.  
\end{equation}
We change variables in order to make the dependence on $1-\gamma^2$ manifest,
\begin{equation}
p_\LCperp^2=(1-\gamma^2)r \qquad p_0=1+(1-\gamma^2)\frac{v}{2} \;.
\label{eq:subsrv}
\end{equation}
Working in an expansion in $1-\gamma^2$, the asymptotic longitudinal momenta read to leading order
\begin{equation}\label{pqg}
p^2=(1-\gamma^2)(v-r) \qquad q^2=(1-\gamma^2)(2-v-r) \;,
\end{equation}
with corrections being of order $\mathcal{O}[(1-\gamma^2)^2]$.  With
this change of variables of \Eqref{eq:subsrv} and working to leading
order in $(1-\gamma^2)$, the effective action becomes
\begin{equation}\label{Gammarv}
\text{Im }\Gamma=\frac{L^2T}{8(2\pi)^2}(1-\gamma^2)^2\int\limits_0^1\ud r\int\limits_r^{2-r}\ud v \;\mathcal{T} \;.
\end{equation}
We can already see that $\text{Im }\Gamma$ vanish at least as fast as
$(1-\gamma^2)^2$, in contrast to the semiclassical exponent
that follows from \Eqref{semiclassical-g}. We will  show that $\mathcal{T}$ is linear in $1-\gamma^2$ for a large
class of fields, so that $\text{Im
}\Gamma\sim(1-\gamma^2)^3$.

For this, we begin by expanding the Klein-Gordon equation \eqref{Klein} to first order in $1-\gamma^2$,
\begin{equation}\label{Klein2}
\left(\partial_u^2-\frac{1-f^2}{k^2}+\frac{1-\gamma^2}{k^2}[f^2+(1-v)f-r]\right)\varphi=0 \;,
\end{equation}
where $u=kx$. In order to find $J$ and $R$ in \Eqref{asymptotic}, we start 
with the asymptotic wave $\varphi=e^{iqu/k}$ for $u\sim k/q\gg1$ and
work backwards to $u\to-\infty$. For definiteness we introduce a
bookkeeping parameter $\lambda\ll1$ and define the asymptotic regions
as
\begin{equation}\label{asymptotic-def}
1-f^2\leq\lambda(1-\gamma^2) \;.
\end{equation} 
For the class of fields defined in \Eqref{1f1}, there are only two
regions satisfying \Eqref{asymptotic-def}, which we refer to as the
right ($f>0$) and the left ($f<0$) asymptotic region. In the right
asymptotic region, the Klein-Gordon equation is simply
\begin{equation}
\left(\partial_u^2+\frac{q^2}{k^2}\right)\varphi=0,
\end{equation}
where we have used \Eqref{pqg} and $f\simeq 1+
  \mathcal{O}(\lambda(1-\gamma^2))$. According to \Eqref{asymptotic},
the solution is $\varphi=e^{iqu/k}$. Let us call the location of
  the border of the right asymptotic region $u_\lambda^\LCp$ defined
by $1-f^2(u_\lambda^\LCp)=\lambda(1-\gamma^2)$ and
$f(u_\lambda^\LCp)>0$. We claim that $\varphi$ is a very slowly
  oscillating wave near $u_\lambda^\LCp$, so that we can choose
  an irrelevant phase such that we have $\varphi=1$ at and
  near the border of the right asymptotic region. This claim follows
trivially for fields that are identically zero for $u$ larger than
some $u_0$, i.e. for fields with $f(u_0)=1$ at $|u_0|<\infty$. For
fields vanishing asymptotically, $u_\lambda^\LCp$ is large so one has
to be more careful to make sure that $qu_\lambda^\LCp/k$ is
small. Using the defining equation for $u_\lambda^\LCp$ and
  assuming that the field decays asymptotically as $E\to
  E_0c(kx)^{-d}$ with $c$ being a dimensionless constant, we have
\begin{equation}\label{qu}
\frac{qu_\lambda^\LCp}{k}=\frac{1}{k}\bigg(\frac{2c}{d-1}\frac{1}{\lambda}\bigg)^\frac{1}{d-1}q^\frac{d-3}{d-1} \;.
\end{equation}   
For fields decaying sufficiently fast, i.e., for $d>3$, the
  plane wave phase \eqref{qu} is indeed small, so
$\varphi=e^{iqu/k}\approx1$ for $u\sim u_\lambda^\LCp$. For $d<3$, on
the other hand, \eqref{qu} is large. Below, we see that weak fields
with $d\leq3$ and strong fields with $d=2$ exhibit different scalings
from $(1-\gamma^2)^3$.

As $u$ decreases from the right asymptotic region we first come to what we will refer to as the right semi-asymptotic region, which is delineated by $f>0$ and
\begin{equation}\label{semi-asymptotic}
\lambda(1-\gamma^2)\leq 1-f^2\leq\Lambda(1-\gamma^2)\ll1 \;,
\end{equation}
where $\Lambda\gg1$ is another bookkeeping parameter. For reference,
we indicate the right semi-asymptotic region by
$u_\Lambda^\LCp<u<u_\lambda^\LCp$. In this region, the second
  and the third term in \eqref{Klein2} can be of the same
order. Since we can still substitute $f=1$ in the third term, the
Klein-Gordon equation \eqref{Klein2} reduces to
\begin{equation}\label{Kleinq}
\left(\partial_u^2-\frac{1-f^2}{k^2}+\frac{q^2}{k^2}\right)\varphi=0 \;.
\end{equation}
However, since both the second and third term in \Eqref{Kleinq} are
small, we can solve for $\varphi$ perturbatively, which to lowest
order is simply $\varphi=1$.

In the left semi-asymptotic region, which satisfies
\Eqref{semi-asymptotic} but with $f<0$, the Klein-Gordon equation is
given by replacing $q^2$ with $p^2$ in \Eqref{Kleinq}. For reference
we indicate this region by $u_\lambda^\LCm<u<u_\Lambda^\LCm$. For
fields vanishing in this region either at a finite point or
asymptotically with $d>3$, the solution to lowest order is given by
$\varphi=C+Du$, where $C$ and $D$ are constants obtained by solving
the Klein-Gordon equation in between the semi-asymptotic
  regions where $1-f^2\gg 1-\gamma^2$. These constants depend on the
microscopic details of the field in the region where the field
  strength is comparatively strong, but they are independent of
$1-\gamma^2$, because $\varphi=1$ in the right semi-asymptotic
region. Since $p|u|/k\ll1$ in the left semi-asymptotic region we can
write
\begin{equation}\label{CD}
\begin{split}
\varphi=&C+Du \\
\simeq&\frac{kD}{2ip}\left\{\exp\left[\frac{ip}{k}\Big(u+\frac{C}{D}\Big)\right]
-\exp\left[-\frac{ip}{k}\Big(u+\frac{C}{D}\Big)\right]\right\} \;.
\end{split}
\end{equation}
By matching the semi-asymptotic form \eqref{CD} with the asymptotic form \eqref{asymptotic} we find
\begin{equation}\label{JfromD}
|J|=\frac{k|D|}{2p}.
\end{equation}
Note that $|R|=|J|$ holds to lowest order as expected, since $|J|\gg1$. With
\Eqref{scalar T} and \Eqref{pqg} the tunneling factor becomes
\begin{equation}\label{T-D}
\mathcal{T}=\frac{4pq}{k^2|D|^2}=\frac{4(1-\gamma^2)}{k^2|D|^2}\sqrt{v-r}\sqrt{2-v-r} \;.
\end{equation}
After substituting this into \Eqref{Gammarv} and performing the
momentum integrals ($v$ and $r$) we finally find
\begin{equation}\label{final3}
\text{Im }\Gamma=\frac{\pi L^2T}{12}\frac{(1-\gamma^2)^3}{E_0^2|2\pi D|^2} \;.
\end{equation}
Since $D$ is independent of $1-\gamma^2$ we see that all fields that
vanish faster than $|x|^{-3}$ have the same power-law scaling
$\text{Im }\Gamma \sim (1-\gamma^2)^\beta$ with critical exponent
$\beta=3$. This scaling can be confirmed by comparing with the exact
result \cite{Nikishov:1970br} for a Sauter field. In order to arrive
at \Eqref{final3}, there was no need to assume $E\ll1$ (or equivalently
$k\ll1$). In fact, in the next sections we will derive explicit
expressions for the non-universal constant $D$ in terms of e.g. $d$
and $n$, both for weak fields $E\ll1$ and for strong fields $E\gg1$.

\begin{figure}
\includegraphics[width=0.45\textwidth]{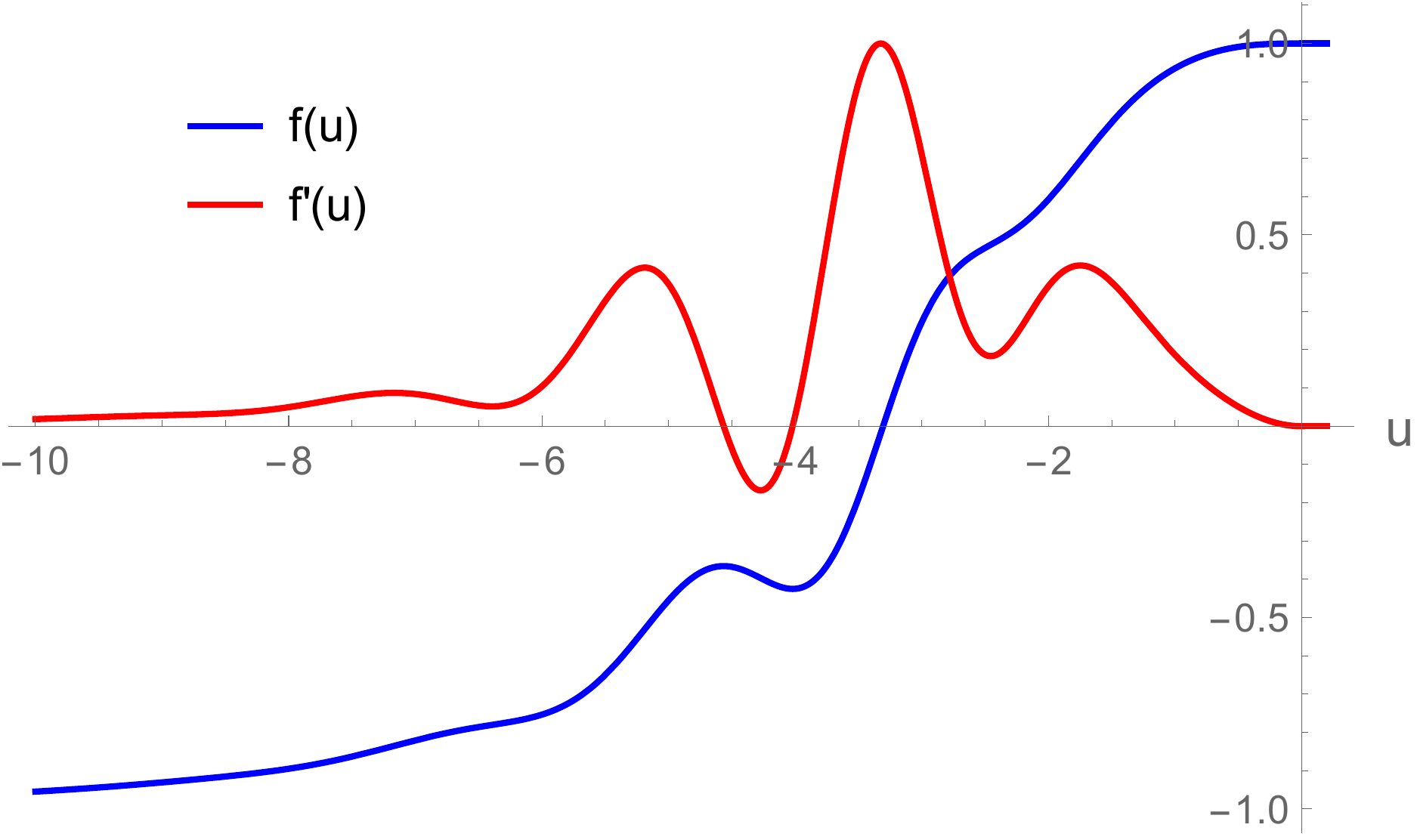}
\caption{Example field as defined in \Eqref{fexample1}. The blue curve depicts the potential
  function $f(kx)$ and the red curve is the electric field
  $E(kx)/E_0=f'$. The electric field has been chosen identically zero
  for $kx>0$ and vanishes asymptotically as $|kx|^{-6}$ for
  $kx\to-\infty$.}
\label{ex1}
\end{figure}

\begin{figure}
\includegraphics[width=0.45\textwidth]{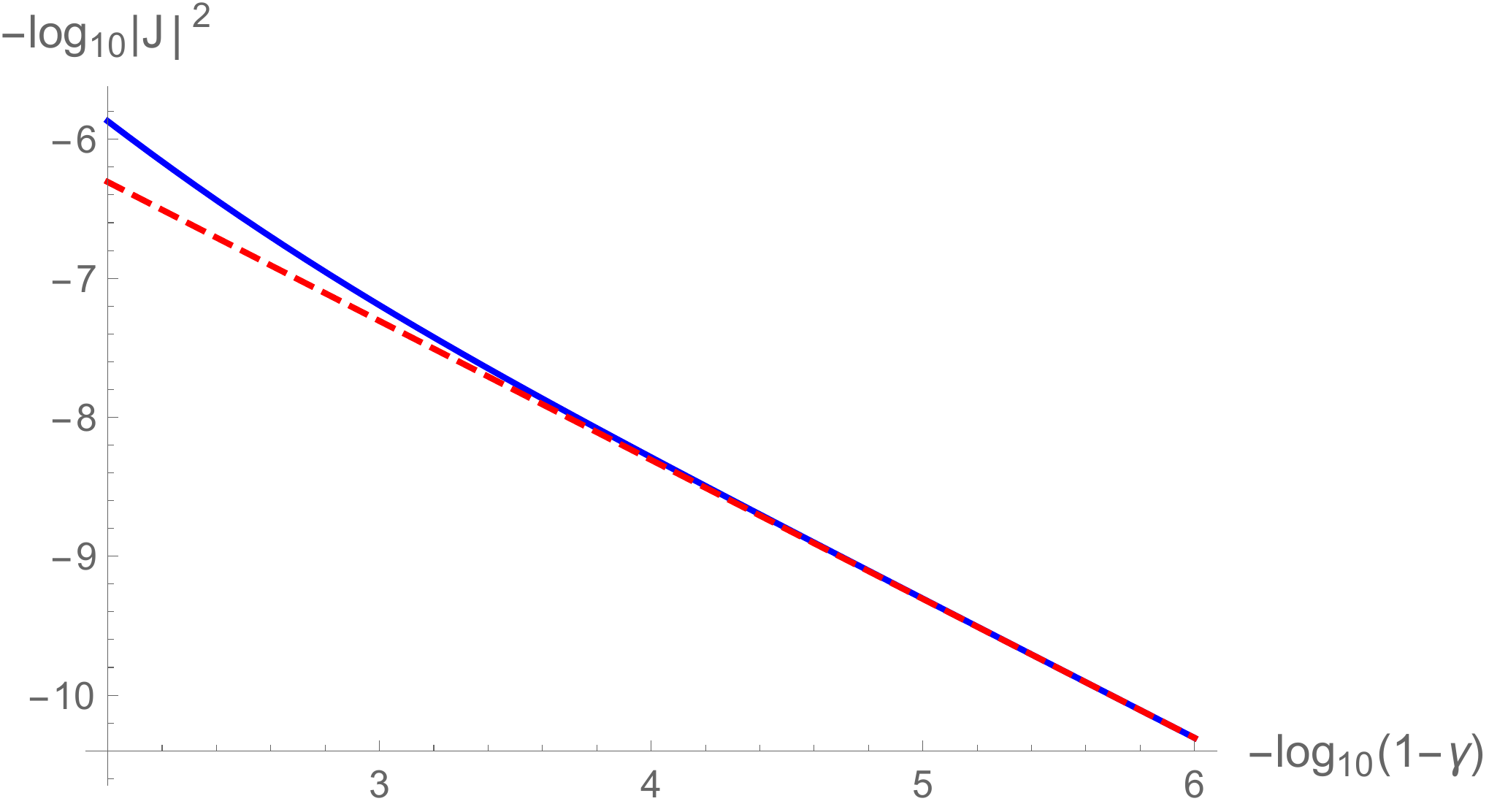}
\caption{Double-logarithmic plot of the inverse of the squared amplitude,
  $1/|J|^2$, as a function of $\gamma$ for the field shown in Fig.~\ref{ex1}. The blue curve is obtained by
  solving \Eqref{Klein} numerically, and the red dashed line is a
  straight line with slope $-1$ obtained by matching with the blue
  curve at the end of the plot. The field strength is $E_0=1$ and
  the momentum parameters at $r=0$ and $v=1$. This plot demonstrates
  that sufficiently close to the critical surface $\gamma=1$ the
  tunneling factor is linear in $1-\gamma^2$ and contributes as such
  to $\text{Im }\Gamma$, which implies that $\text{Im
  }\Gamma\propto(1-\gamma^2)^3$.}
\label{scaling1}
\end{figure}

\begin{figure}
\includegraphics[width=0.45\textwidth]{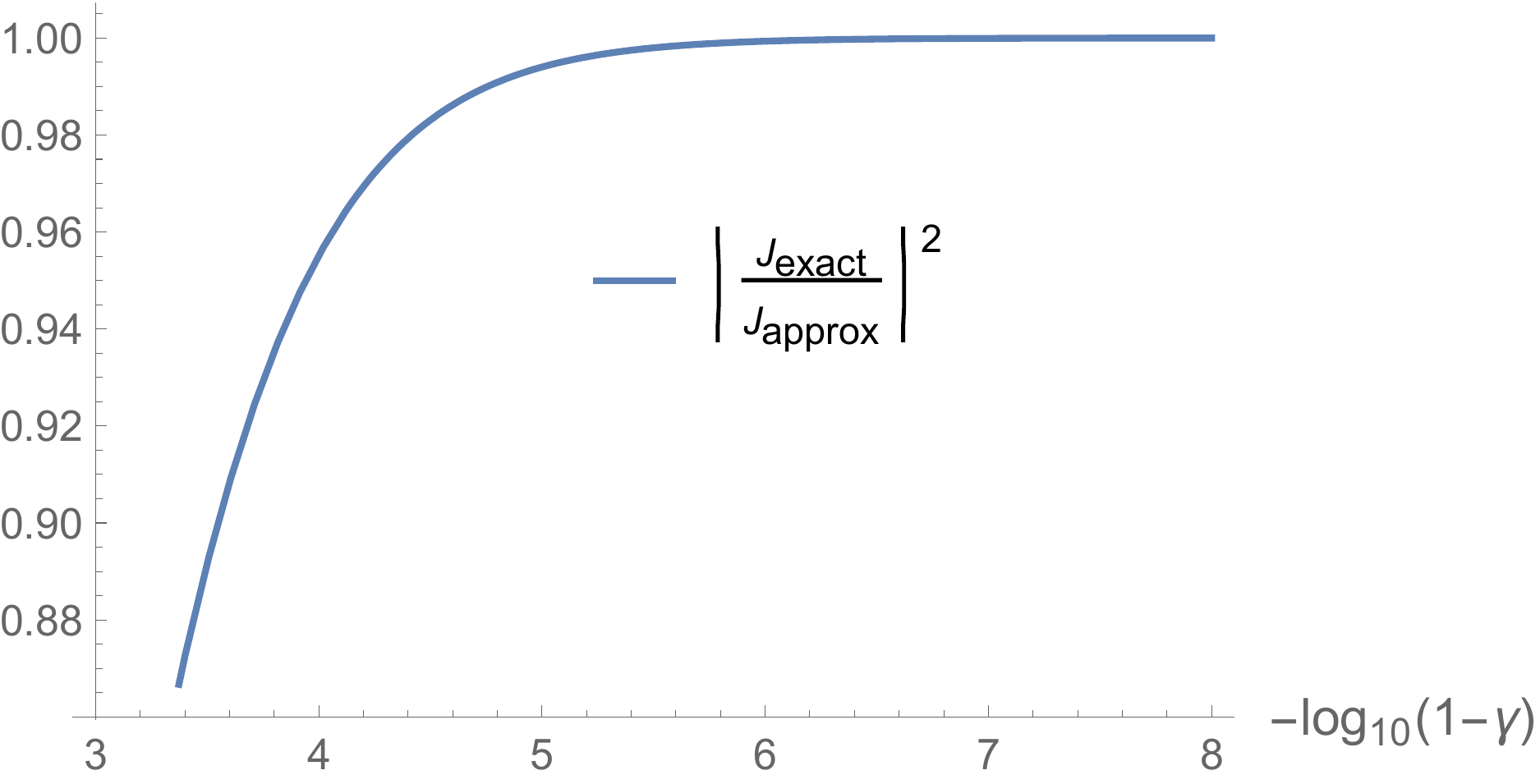}
\caption{Amplitude ratio for the field as in Fig.~\ref{scaling1} with the same parameters. $J_\text{exact}$ is the exact amplitude obtained from the numerical solution of \Eqref{Klein}, and $J_\text{approx}=(E_0/2p)\varphi'(u\to-\infty)$ is obtained from the numerical solution of \Eqref{Klein3} with boundary condition $\varphi=1$ and $\varphi'=0$ at $u=0$. For $\gamma=1-10^{-8}$ we have $|J_\text{exact}/J_\text{approx}|^2\approx1-7*10^{-6}$.}
\label{exact-critical-approx-ex1-fig}
\end{figure}

As a verification of \Eqref{final3}, we have numerically solved the
Klein-Gordon equation in its original form \eqref{Klein}. The result
for the field in Fig.~\ref{ex1} is shown in
Fig.~\ref{scaling1} and Fig.~\ref{exact-critical-approx-ex1-fig}. We summarize the parametrizations of the
  field profiles used as illustrations in the appendix.

\section{D for strong fields}\label{D for strong fields}

Let us now derive the non-universal constant $D$, starting with the
simpler case for strong fields $E_0\gg1$, the weak field case
$E_0\ll1$ is treated in the next section.

For this, we need to solve the equation
\begin{equation}\label{Klein3}
\left(\partial_u^2-\frac{1-f^2}{k^2}\right)\varphi=0
\end{equation}
in between the two semi-asymptotic regions, i.e., for
$u_\Lambda^\LCm<u<u_\Lambda^\LCp$, and with boundary condition
$\varphi=1$ for $u\sim u_\Lambda^\LCp$. Schematically, we need to find
D in the solution chain
\begin{equation}
Je^{ipx}+Re^{-ipx}\,\overset{u_\lambda^\LCm}{\leftarrow}\,C+Du\,\overset{u_\Lambda^\LCm}{\leftarrow}\,\text{?}\,\overset{u_\Lambda^\LCp}{\leftarrow}\, 1\,\overset{u_\lambda^\LCp}{\leftarrow} e^{iqx} \;.
\end{equation}  
The solution to \Eqref{Klein3} is obtained by expanding in $1/k^2$, which to lowest order gives
\begin{equation}
\varphi=1+\int\limits_{u_\Lambda^\LCp}^u\ud u'\; (u-u')\frac{1-f^2(u')}{k^2}, \quad u_\Lambda^\LCm<u<u_\Lambda^\LCp\;.
\end{equation}
By taking $u\sim u_\Lambda^\LCm$ we find 
\begin{equation}\label{D-strong}
D=\int\limits_{-\infty}^\infty\ud u\; \frac{1-f^2(u)}{k^2} \;,
\end{equation}
where the extension of the integration boundaries remains exact to
leading order where $f\simeq1$ in all (semi-)asymptotic
regions. Substituting \Eqref{D-strong} into \Eqref{final3} gives us
\Eqref{final-strong}.

\begin{figure}
\includegraphics[width=0.45\textwidth]{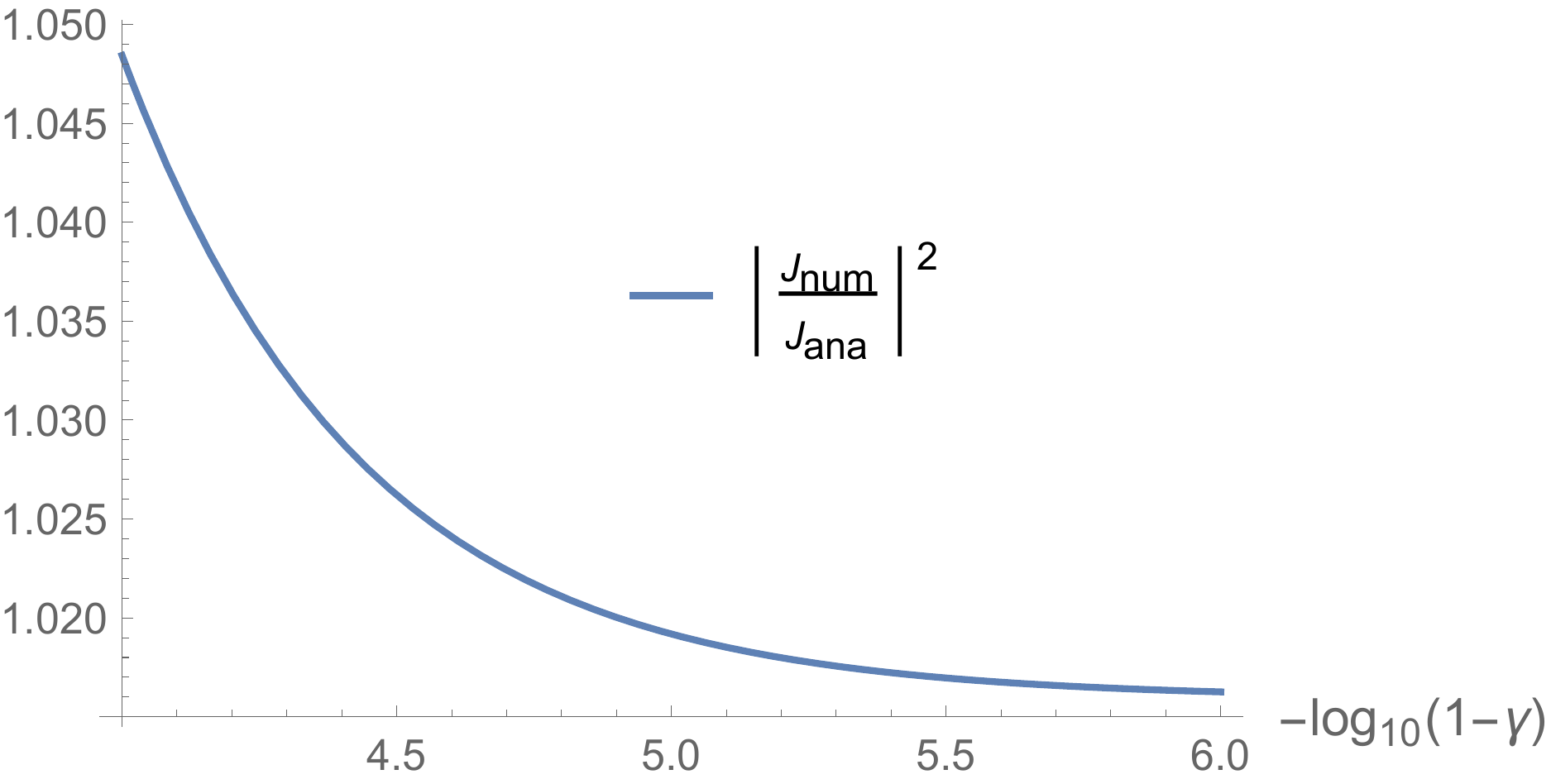}
\caption{Ratio of the squared amplitude $|J|^2$ obtained numerically
  and that obtained analytically from \Eqref{D-strong}, as a function
  of $-\log_{10}(1-\gamma)$. The field shape is that shown in
  Fig.~\ref{ex1}, the field strength is $E_0=30$, and the momentum
  parameters are $r=0$ and $v=1$. The plot shows that the
    analytical approximation for the tunneling factor
    $\mathcal{T}\propto1/|J|^2$ is only slightly larger than the
    numerical result, and that the analytical approximation improves
    with $\gamma\to1$. The asymptotic error can be made smaller by
    choosing a stronger field (recall that \Eqref{D-strong} has been
    derived under the assumptions $1-\gamma^2\ll1/E^2\ll1$).}
\label{strong1}
\end{figure}

\begin{figure}
\includegraphics[width=0.45\textwidth]{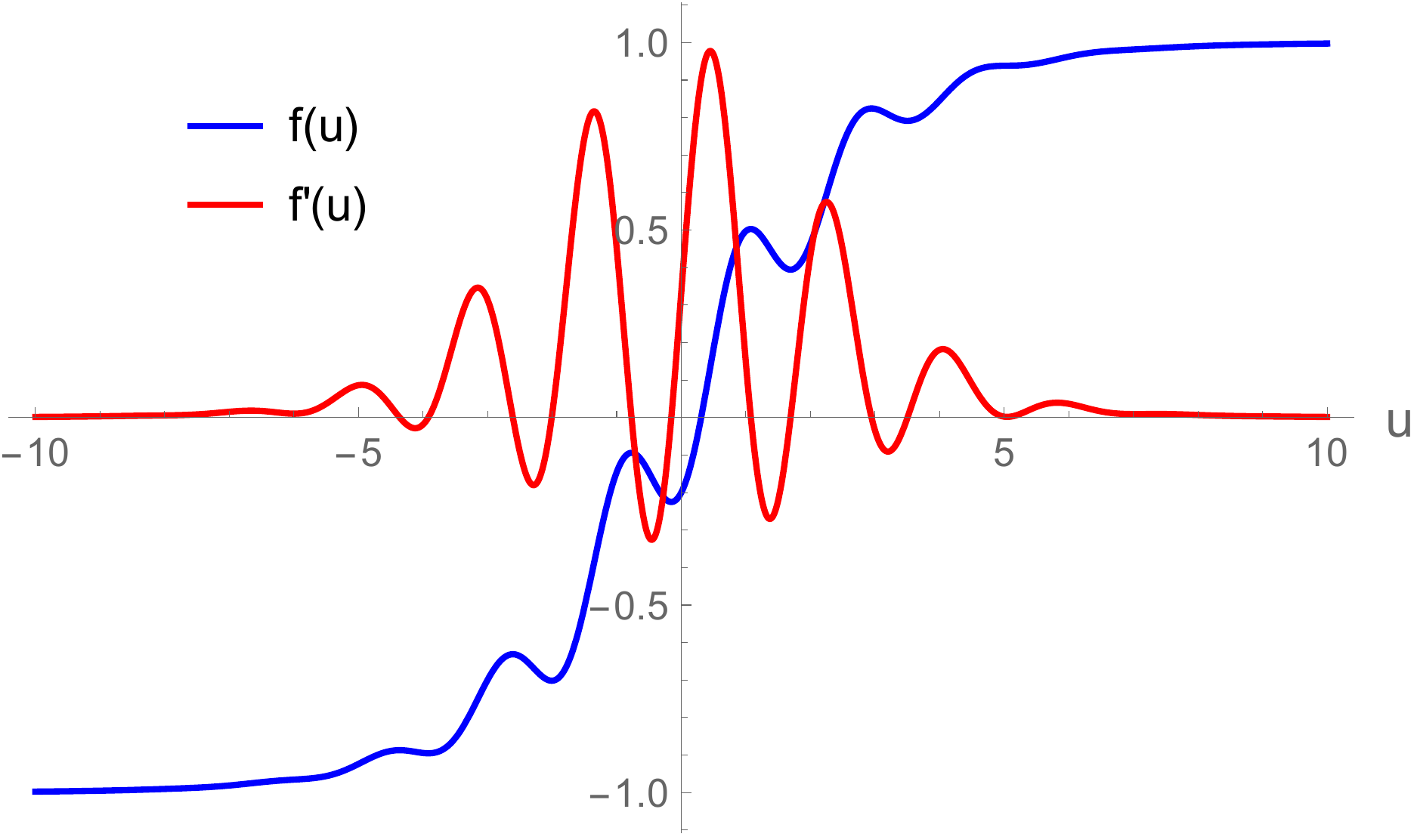}
\caption{Example field as defined in \Eqref{fexample2}. The blue curve
  depicts the potential function $f(kx)$ and the red curve is the
  electric field $E(kx)/E_0=f'$. The field decays exponentially for
  $|u|\to\infty$.}
\label{ex2}
\end{figure}

\begin{figure}
\includegraphics[width=0.45\textwidth]{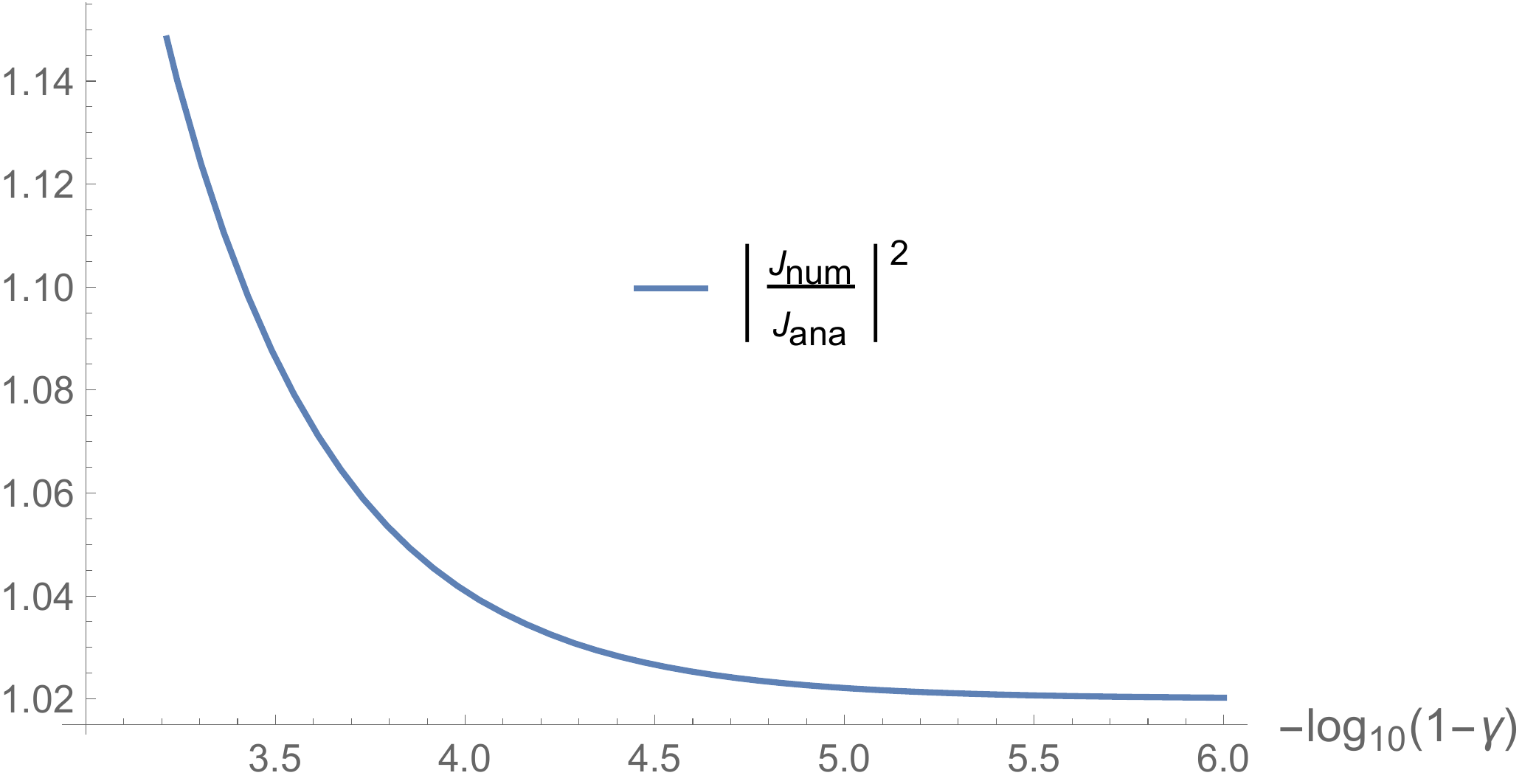}
\caption{Ratio of the squared amplitude $|J|^2$ obtained numerically
  and that obtained analytically from \Eqref{D-strong}, as a function
  of $-\log_{10}(1-\gamma)$. The field shape is that shown in
  Fig.~\ref{ex2}, the field strength is $E_0=30$, and the momentum
  parameters are $r=0$ and $v=1$. The plot shows that the analytical
  approximation improves with $\gamma\to1$.}
\label{strong2}
\end{figure}

\begin{figure}
\includegraphics[width=0.45\textwidth]{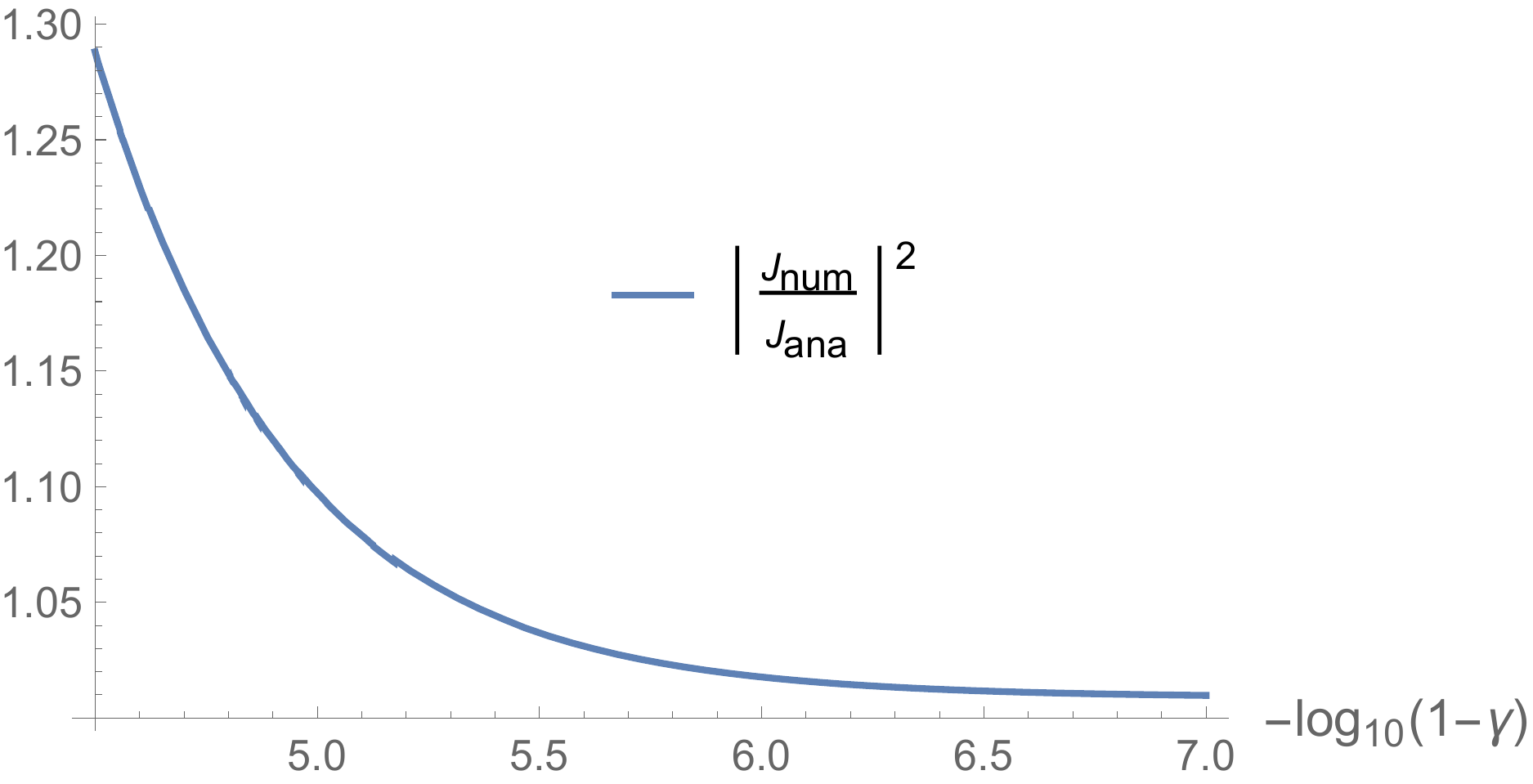}
\caption{Ratio of the squared amplitude $|J|^2$ obtained numerically
  and that obtained analytically from \Eqref{D-strong}, as a function
  of $-\log_{10}(1-\gamma)$. The field shape is
  $f=u(1+|u|^{3/2})^{-2/3}$, with field-strength parameter $E_0=200$,
  and momentum parameters $r=0$ and $v=1$. The plot shows that the
  strong-field approximation \eqref{final-strong} is valid also for
  fields vanishing slower than $|x|^{-3}$. However, here we need
  larger $E_0$ for \Eqref{D-strong} to be a satisfactory
  approximation. This serves as an indication for the fact that we
  find a completely different scaling for weak fields vanishing slower
  than $|x|^{-3}$.}
\label{fig-strong-slow}
\end{figure}

As a simple check of \Eqref{D-strong}, consider a Sauter pulse,
$f(u)=\tanh u$. From \Eqref{D-strong} and \Eqref{T-D} it follows that
$\mathcal{T}=qp k^2$, which is in perfect agreement with the exact
solution \cite{Nikishov:1970br,Holstein:1999ta} in the regime
$1-\gamma^2\ll1/E^2\ll1$.  As a further check, consider the field
depicted in Fig.~\ref{ex1}. In Fig.~\ref{strong1} we show that
\Eqref{D-strong} agrees well with a numerical solution of the
Klein-Gordon equation \eqref{Klein}. Another field example is shown in
Fig.~\ref{ex2}, and the agreement between \Eqref{D-strong} and
numerical results is demonstrated in Fig.~\ref{strong2}. In
Fig.~\ref{fig-strong-slow} we demonstrate with a field that vanishes
as $|x|^{-5/2}$ that \Eqref{D-strong} also holds for fields vanishing
slower than $|x|^{-3}$ (compare though with the scaling found in
Sect.~\ref{Fields decaying as x2} for fields vanishing as $|x|^{-2}$).
In each of these cases, the asymptotic error for $\gamma\to1$,
  i.e., the slight deviations of the numerical to analytical amplitude
  ratio from the deeply critical value
  $|J_{\text{num}}/J_{\text{ana}}|^2 \to 1$, is controlled by the
  field strength. The deeply critical amplitude ratio is approached
  more closely, the better the parameter hierarchy
  $1-\gamma^2\ll1/E^2\ll1$ of the deeply critical regime is
  satisfied. We emphasize that this asymptotic error does not affect
  the scaling property.
     
\section{D for weak fields}\label{D for weak fields}

Let us now derive the non-universal coefficient $D$ in \Eqref{final3}
for weak fields $E_0\ll1$. We will consider
fields that vanish either asymptotically as $x^{-d}$ or
beyond specific points $x_0$ as $(x-x_0)^n$; there are four
different combinations.

We again have to solve \Eqref{Klein3} in the region between the two
semi-asymptotic regions, $u_\Lambda^\LCm<u<u_\Lambda^\LCp$, and with
boundary condition $\varphi=1$ for $u\sim u_\Lambda^\LCp$. This time
we divide this region into three regions. We begin with the two
outer regions where
\begin{equation}
1-f^2\leq L k^2\ll1 \qquad L\gg1 \;,
\end{equation}
where $L$ is yet another bookkeeping parameter. In these two regions
we expand $f$ around $\pm1$. The lowest nontrivial order can be
  solved analytically in terms of Bessel functions. Therefore, we
refer to these two regions as the left and right Bessel regions from
now on. Defining the inner boundaries of these regions in terms
  of $u_L^\LCpm$ by $1-f(u_L^\LCpm)^2=L k^2$, where the $\pm$ sign
  holds for $f>0$ and $f<0$, respectively, the left Bessel region
covers $u_\Lambda^\LCm<u<u_L^\LCm$ and the right Bessel region
$u_L^\LCp<u<u_\Lambda^\LCp$.

Between the two Bessel regions, i.e. $u_L^\LCm<u<u_L^\LCp$, we have
\begin{equation}
\frac{1-f^2}{k^2}\geq L\gg1 \;,
\end{equation} 
which implies that \Eqref{Klein3} can be well approximated by
the standard WKB method,
\begin{equation}\label{WKB}
\varphi=\frac{A}{(1-f^2)^\frac{1}{4}}\exp\left(\int\limits_u^\infty\frac{\sqrt{1-f^2}}{k}\right) \;.
\end{equation}
We have dropped the term with opposite sign in front of the integral
since it is exponentially smaller. The upper limit in the integral is
chosen for convenience. This choice
is possible for all $n$ and for $d>3$, but for $d\le3$ we have to
choose a finite value.

To connect the WKB and the Bessel regions we note that near the
boundaries $u\sim u_L^\LCpm$ we can express the solution in terms of
either Bessel functions or as in \Eqref{WKB}. In other words, the
region where the WKB form is valid partly overlap with the region
where we can expand $f$ around $\pm1$. By expanding the Bessel
functions in $u$ near the boundary between the left Bessel region and
the left semi-asymptotic region, i.e. at $u\sim u_\Lambda^\LCm$, we
obtain $D$ in \Eqref{CD},  which then completes the final result
\Eqref{final3}.

In summary, we wish to construct the solution chain for weak fields,
\begin{equation}\label{summary}
\begin{split}
Je^{ipx}+Re^{-ipx}\,\overset{u_\lambda^\LCm}{\leftarrow} &\, C+Du\,\overset{u_\Lambda^\LCm}{\leftarrow}\text{Bessel}\,\leftarrow \\ 
\overset{u_L^\LCm}{\leftarrow} &\,\text{WKB}\,\overset{u_L^\LCp}{\leftarrow}\,\text{Bessel}\,\overset{u_\Lambda^\LCp}{\leftarrow}\, 1\,\overset{u_\lambda^\LCp}{\leftarrow}\, e^{iqx} \;. 
\end{split}
\end{equation}  

\subsection{The right Bessel region}

\subsubsection{Fields decaying asymptotically}

We begin with the right Bessel region, $u_L^\LCp<u<u_\Lambda^\LCp$,
and with fields decaying as
\begin{equation}\label{asymptoticE}
E\to E_0\frac{c_\LCp}{(kx)^{d_\LCp}} \qquad x\to\infty\;,
\end{equation}
where $d_\LCp>3$. The $+$ subscripts indicate that we are in the right
Bessel region; to avoid cumbersome notation we will simply write $d$
and $c$ where the meaning should be clear from the context. The
potential is
\begin{equation}
f=1-\frac{c}{d-1}u^{-(d-1)},
\end{equation}
and the Klein-Gordon equation reduces to
\begin{equation}
\Big(\partial_u^2-\frac{2c}{(d-1)k^2}u^{-(d-1)}\Big)\varphi=0 \;.
\end{equation}
The solution can be written in terms of Bessel functions
\cite{NIST:DLMF}. The correct linear combination is determined by
requiring that $\varphi\to1$ as $u\to\infty$
(c.f. \Eqref{summary}). One obtains
\begin{equation}\label{Besselp}
\varphi=b_1\sqrt{u}I\bigg(\frac{1}{d-3},\frac{2}{d-3}\sqrt{\frac{2c}{d-1}}\frac{1}{ku^\frac{d-3}{2}}\bigg) \;,
\end{equation}
with normalization constant
\begin{equation}
b_1=\Gamma\bigg(1+\frac{1}{d-3}\bigg)\bigg(k(d-3)\sqrt{\frac{d-1}{2c}}\bigg)^\frac{1}{d-3} \;.
\end{equation}
By matching \Eqref{Besselp} and the WKB form \eqref{WKB} for $u\sim u_L^\LCp$,
we find \cite{NIST:DLMF}
\begin{equation}
A=\sqrt{\frac{k(d-3)}{4\pi}}b_1 \;.
\label{eq:wfas}
\end{equation}

\subsubsection{Fields vanishing beyond a finite point}

Before we connect with the left Bessel region we consider fields vanishing beyond a finite point $u_\LCp$ as
\begin{equation}
E\to E_0c_\LCp(u_\LCp-u)^{n_\LCp}\;.
\end{equation}   
The potential is given by (again omitting $+$ subscripts for brevity)
\begin{equation}
f=1-\frac{c}{n+1}(u_\LCp-u)^{n+1}, \quad \text{for}\,\, u\lesssim u_\LCp \;,
\end{equation}
and $f=1$ for $u\geq u_{\LCp}$. The Klein-Gordon equation
reduces to 
\begin{equation}
\Big(\partial_{v_\LCp}^2-\frac{2c}{(n+1)k^2}v_\LCp^{n+1}\Big)\varphi=0 \;,
\end{equation}
where $v_\LCp=u_\LCp-u$. The general solution is again given by
Bessel functions. Demanding that the solution and its
  derivative be continuous at $u_\LCp$, we find
\begin{equation}\label{besseln}
\varphi
=b_1\sqrt{v_\LCp}I\bigg(-\frac{1}{3+n},\frac{2}{3+n}\sqrt{\frac{2c}{1+n}}\frac{v_\LCp^\frac{3+n}{2}}{k}\bigg) \;,
\end{equation}
where
\begin{equation}
b_1=\Gamma\bigg(1-\frac{1}{3+n}\bigg)\bigg(\frac{1}{3+n}\sqrt{\frac{2c}{1+n}}\frac{1}{k}\bigg)^\frac{1}{3+n} \;.
\end{equation}
By matching \Eqref{besseln} with the WKB form \eqref{WKB} in the
overlap region $u\sim u_L^\LCp$,
we find \cite{NIST:DLMF}
\begin{equation}
A=\sqrt{\frac{k(3+n)}{4\pi}}b_1 \;.
\end{equation}
Note that the similarity between the asymptotic ($d$) case in
\Eqref{eq:wfas} and the compact ($n$) case here. To highlight this, we
let $s$ stand for either $n$ or $-d$, and define
\begin{equation}\label{F}
F(s,c):=|3+s|^\frac{1}{2}\Gamma\bigg(1-\frac{1}{3+s}\bigg)\bigg(\frac{1}{|3+s|}\sqrt{\frac{2c}{|1+s|}}\frac{1}{k}\bigg)^\frac{1}{3+s} \;.
\end{equation}
This allows us to write the WKB constant in \Eqref{WKB} as
\begin{equation}\label{AF}
A(s,c)=\sqrt{\frac{k}{4\pi}}F(s,c) \;.
\end{equation}

\subsection{The left Bessel region}

Next we perform a similar matching with the Bessel region to the left
of the WKB region at $u_L^\LCm$.

\subsubsection{Fields vanishing asymptotically}

We begin again with fields decaying asymptotically,
\begin{equation}
E\to E_0\frac{c_\LCm}{(-kx)^{d_\LCm}} \;,
\end{equation}
where $d_\LCm>3$. Note that the constants $c_\LCm,d_\LCm$
  specifying the asymptotics of the field profile are allowed to be
  different from those for the $x\to+\infty$ asymptotics. Dropping
  again the subscripts for brevity, the potential function is
\begin{equation}
f=-1+\frac{c}{d-1}(-u)^{-(d-1)},
\end{equation}
and the Klein-Gordon equation reduces to
\begin{equation}
\Big(\partial_u^2-\frac{2c}{(d-1)}\frac{1}{k^2(-u)^{(d-1)}}\Big)\varphi=0 \;.
\end{equation}
By matching with the WKB form in the overlap region near
  $u_L^\LCm$, we find for the left Bessel region
$u_\Lambda^\LCm<u<u_L^\LCm$
\begin{equation}\label{bessel-left-d}
\varphi=b_3\sqrt{-u}K\bigg(\frac{1}{d-3},\frac{2}{d-3}\sqrt{\frac{2c}{d-1}}\frac{1}{k(-u)^\frac{d-3}{2}}\bigg) \;,
\end{equation}
where
\begin{equation}
b_3=\frac{e^S}{\pi}\frac{F(s_\LCp,c_\LCp)}{\sqrt{d_\LCm-3}}.
\end{equation}
Here, the tunnel exponent $S$, as in \Eqref{Sint}, arises from the
integral in the WKB exponent \Eqref{WKB}.

\subsubsection{Fields vanishing beyond a finite point}

Finally, we consider fields approaching zero at $u\to u_\LCm$ as
\begin{equation}
E\to E_0c_\LCm(u-u_\LCm)^{n_\LCm}\;,
\end{equation}   
and vanishing beyond that point for all $u<u_\LCm$. The potential
is given by
\begin{equation}
f=-1+\frac{c}{n+1}(u-u_\LCm)^{n+1}, \quad \text{for}\,\, u\gtrsim u_\LCm \;,
\end{equation}
and $f=-1$ for $u\leq u_{\LCp}$.  The Klein-Gordon equation
reduces to
\begin{equation}
\Big(\partial_{v_\LCm}^2-\frac{2c}{(n+1)k^2}v_{\LCm}^{n+1}\Big)\varphi=0 \;,
\end{equation}
where $v_\LCm=u-u_\LCm$. A similar matching as above gives us the solution in the left Bessel region, 
\begin{equation}\label{bessel-left-n}
\varphi=b_3\sqrt{v_\LCm}K\bigg(\frac{1}{3+n},\frac{2}{3+n}\sqrt{\frac{2c}{1+n}}\frac{v_\LCm^\frac{3+n}{2}}{k}\bigg) \;,
\end{equation}
where
\begin{equation}
b_3=\frac{e^S}{\pi}\frac{F(s_\LCp,c_\LCp)}{\sqrt{3+n_\LCm}} \;.
\end{equation}
For fields vanishing at either $u_\LCp$ or $u_\LCm$ or both, it is
obvious that the integrand in $S$ given by \Eqref{Sint} is
identically zero beyond these points.

\subsection{The left asymptotic region and the final result for $D$}

We are now in a position to obtain $D$ in \Eqref{final3}. For fields
decaying asymptotically in the left Bessel region, we take the
$u\to-\infty$ limit of \Eqref{bessel-left-d} and compare with
\Eqref{CD}, and for fields vanishing at a finite point in the left
Bessel region we take the $v\to0$ limit of \Eqref{bessel-left-n}.  In
all cases we find
\begin{equation}\label{DFF}
|D|=\frac{e^S}{2\pi}F(s_\LCm,c_\LCm)F(s_\LCp,c_\LCp) \;,
\end{equation}
where $s=n$ or $s=-d$ depending on the asymptotic field properties,
and $F$ as in \Eqref{F}.
This is the final result valid for weak fields $E_0\ll1$ and for all $n$ and $d>3$.

\begin{figure}
\includegraphics[width=0.45\textwidth]{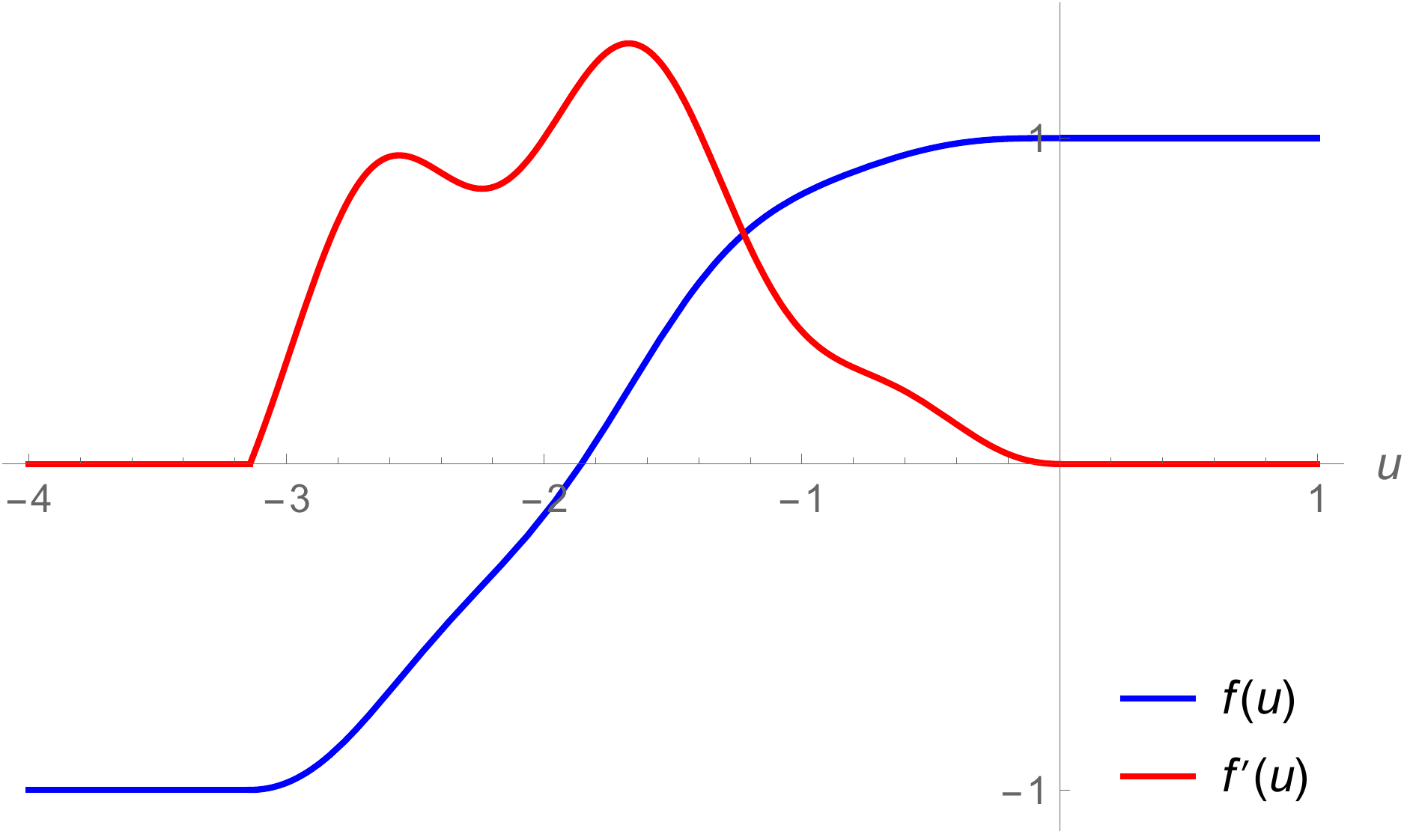}
\caption{Compact field example, defined by~\eqref{fCompactEx}. The blue curve is the potential $f(kx)$ and the red curve is the electric field $E(kx)/E_0=f'$. The electric field is identically zero for $kx>0$ and $kx<-\pi$.}
\label{nontrivial-weak-compact-fig}
\end{figure}

\begin{figure}
\includegraphics[width=0.45\textwidth]{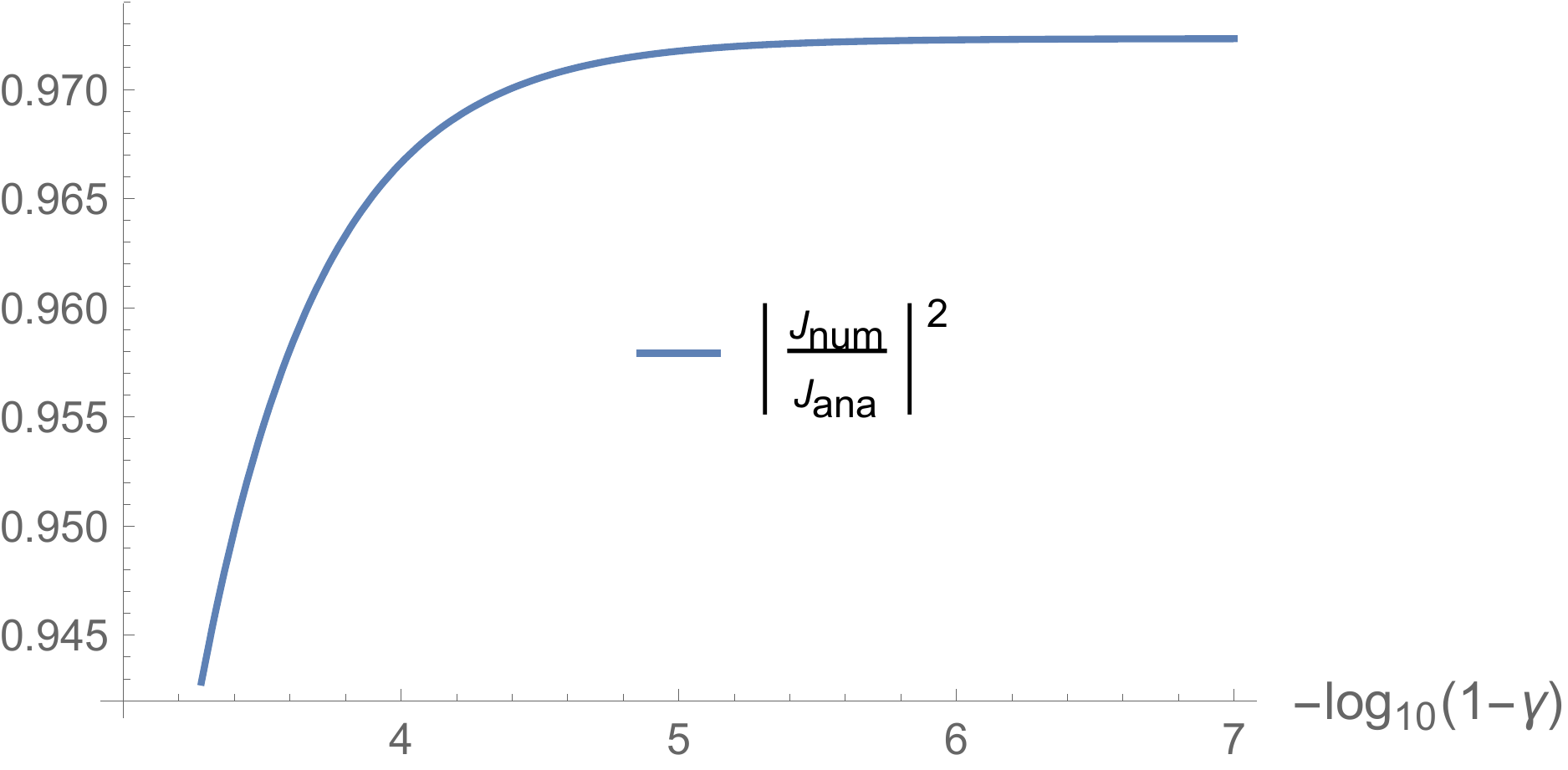}
\caption{Ratio of the squared amplitude $|J|^2$ obtained numerically
  and analytically. The field shape is that shown in
  Fig.~\ref{nontrivial-weak-compact-fig}, the field strength is
  $E_0=0.07$, and the momentum parameters are $r=0$ and $v=1$. The
  plot shows that the analytical approximation becomes better as
  $\gamma\to1$.}
\label{nontrivial-weak-compact-amplitude-fig}
\end{figure}

\begin{figure}
\includegraphics[width=0.45\textwidth]{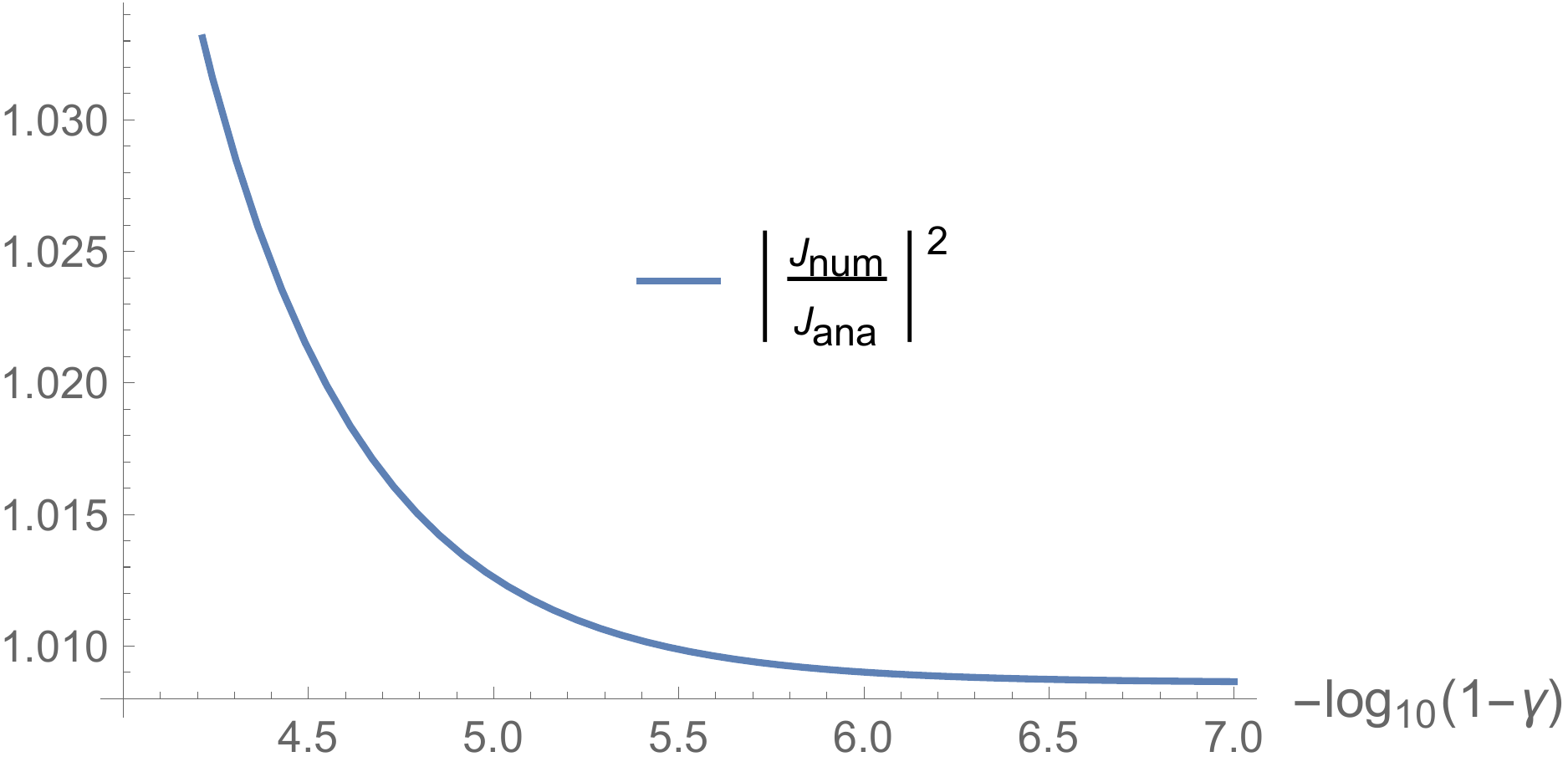}
\caption{Ratio of $|J|^2$ obtained numerically and analytically using \Eqref{D-strong}. The field shape is that shown in Fig.~\ref{nontrivial-weak-compact-fig}, the field strength is $E_0=10$, and the momentum parameters are $r=0$ and $v=1$. The plot shows that the analytical approximation becomes better as $\gamma\to1$.}
\label{nontrivial-strong-compact-amplitude-fig}
\end{figure}

To check \Eqref{DFF}, we solve \Eqref{Klein} numerically for the
electric field shown in Fig.~\ref{nontrivial-weak-compact-fig}. The
ratio of the amplitude squared obtained analytically and numerically
is shown in Fig.~\ref{nontrivial-weak-compact-amplitude-fig}. (See
Fig.~\ref{nontrivial-strong-compact-amplitude-fig} for the
corresponding plot in the strong field regime.) For some fields,
  a quantitative comparison requires to choose $E_0$ rather small
depending on the desired precision of $D$, because the regions where
it is sufficient to keep the first order corrections to $f=\pm1$ have
to partly overlap with the WKB region; due to the exponential suppression, this can make the corresponding $\text{Im }\Gamma$ very small.
We emphasize, though, that this does not affect the
scaling.

As a further check, we consider the limit $d\to\infty$. From our
previous paper, we expect this to give the same result as an
exponentially decaying field, in particular as the exact result for
the Sauter field \cite{Nikishov:1970br}. Writing
\begin{equation}
e^{-\kappa u}=\lim\limits_{d\to\infty} \Big(1+\frac{\kappa u}{d}\Big)^{-d}
\end{equation} 
and casually exchanging the order of the limits $d\to\infty$ and
$u\to\infty$, suggests that we should scale $c$ in \Eqref{asymptoticE}
as $c=(d/\kappa)^d\bar{c}$ with some $d$-independent constant
$\bar{c}$. With this rescaling, the $d\to\infty$ limit of \Eqref{DFF}
is $D=\kappa e^S/2\pi$, and the imaginary part of the effective action
becomes
\begin{equation}\label{Gamma exp}
\text{Im }\Gamma=\frac{\pi}{12}L^2T\frac{(1-\gamma^2)^3}{\kappa^2 E^2}e^{-2S} \;.
\end{equation}
For the Sauter field, we have $\kappa=2$ and $S=\pi/E$ from
\Eqref{Sint}, which indeed agrees with the exact result
\cite{Nikishov:1970br}.

We also recover this result from a limit of compact fields.
In order to compare with an exponentially decaying field $E\to E_0\bar{c}e^{-\kappa u}$ we choose $u_\LCpm=\pm n/\kappa$ and $c=\bar{c}(\kappa/n)^n$, so
\begin{equation}
c(u_\LCp-u)^n=\bar{c}\Big(1-\frac{\kappa u}{n}\Big)^n \;
\end{equation}
and similarly for $u_\LCm$. In the limit $n\to\infty$ we again find
$D=\kappa e^S/2\pi$ and thus \Eqref{Gamma exp}.

\subsection{Exponentially decaying fields}

We have just shown that the result for exponentially decaying fields
can be obtained from limits of fields vanishing with a power either
asymptotically or beyond finite points. For completeness, we derive
the same results directly in this subsection by starting with fields
with asymptotic behavior
\begin{equation}
E\to E_0c\kappa e^{-\kappa|u|} \;.
\end{equation}
Here, $c$ and $\kappa$ may be chosen differently for $x\to\pm\infty$ in
order to allow for fields vanishing non-symmetrically. For $u\gg1$ we
have
\begin{equation}
f=1-ce^{-\kappa u} \;.
\end{equation}
For such fields, we can solve the Klein-Gordon equation in the Bessel,
the semi-asymptotic and the asymptotic regions in one fell swoop. The
solution to
\begin{equation}
\Big(\partial_u^2+\frac{1}{k^2}\Big[-2ce^{-\kappa u}+q^2\Big]\Big)\varphi=0
\end{equation}
can be written as
\begin{equation}
\varphi=\Gamma\bigg(1-\frac{2iq}{\kappa k}\bigg)\bigg(\frac{2c}{\kappa^2k^2}\bigg)^\frac{iq}{\kappa k}I\bigg(-\frac{2iq}{\kappa k},\frac{2\sqrt{2c}}{\kappa k}e^{-\frac{\kappa u}{2}}\bigg) \;.
\end{equation}
By matching with the WKB form \eqref{WKB}, we find
\begin{equation}
A=\sqrt{\frac{\kappa k}{4\pi}} \;.
\end{equation}
This agrees with the $n\to\infty$ and $d\to\infty$ limits of
\Eqref{AF}. By a similar matching in the left Bessel region we find
\begin{equation}
D=\frac{\sqrt{\kappa_\LCm \kappa_\LCp}e^S}{2\pi} \;,
\end{equation}
which agrees with \Eqref{Gamma exp} in the simplifying limit $\kappa_\LCm =\kappa=\kappa_\LCp$.

\section{Weak fields decaying as $E\sim x^{-3}$}\label{Fields decaying as x3}

All the fields considered so far, both compact and asymptotically
decaying with power $d>3$, have the same universal power-law scaling
with critical exponent $\beta=3$. In this section, we consider
weak fields $E_0\ll1$ for the special case $d=3$, and for simplicity
we assume a symmetric decay
\begin{equation}
E\to E_0\frac{c}{|kx|^3} \qquad x\to\pm\infty \;.
\end{equation}
In our previous paper \cite{Gies:2015hia}, we found that these fields
have power-law scaling with a critical exponent depending on the field
strength in the semiclassical regime. We show here that the same
scaling is recovered also in the deeply critical regime.

For $u\gg1$ we have
\begin{equation}
f=1-\frac{c}{2u^2} \;.
\end{equation}
Because the integrand in $S$ goes like $1/u$ we cannot choose the
upper integration limit as in \Eqref{WKB}. Instead we take
\begin{equation}\label{WKB3}
\varphi=\frac{A}{(1-f^2)^\frac{1}{4}}\exp\int\limits_u^U\frac{\sqrt{1-f^2}}{k} \;,
\end{equation} 
with $U$ large but finite. The limit $U\to\infty$ is considered
  on the level of the final result for $\text{Im }\Gamma$. The
solution to
\begin{equation}
\Big(\partial_u^2+\frac{1}{k^2}\Big[-\frac{c}{u^2}+q^2\Big]\Big)\varphi=0
\end{equation}
is given in terms of a Hankel function (Bessel function of the third kind) \cite{NIST:DLMF}
\begin{equation}\label{hankelq}
\varphi=\sqrt{\frac{\pi q u}{2k}}e^{\frac{i\pi}{4}(1+\frac{2\sqrt{c}}{k})}H^{(1)}\Big(\frac{\sqrt{c}}{k},\frac{q u}{k}\Big) \;,
\end{equation}
where the normalization coefficient is chosen such that $\varphi\to
e^{iqx}$ asymptotically. We have neglected some factors that are small
due to $k\ll1$, as we are working in the critical regime
  $\gamma=k/E_0\to1$ and at weak fields $E_0\ll1$. In the overlap
with the WKB region, we have
\begin{equation}
\varphi=Ac^{-\frac{1}{4}}u^{\frac{1}{2}-\frac{\sqrt{c}}{k}}e^{\frac{\sqrt{c}}{k}\ln U} \;,
\end{equation}
which, upon comparing with \Eqref{hankelq}, gives us $A$. In the
overlap between the WKB and the left Bessel region, the solution is
\begin{equation}
\varphi=Ac^{-\frac{1}{4}}(-u)^{\frac{1}{2}+\frac{\sqrt{c}}{k}}e^{S_\Lambda-\frac{\sqrt{c}}{k}\ln U} \;,
\end{equation}
where 
\begin{equation}
S_U=\int\limits_{-U}^U\!\ud u\;\frac{\sqrt{1-f^2}}{k} \;.
\end{equation}
It follows that in the left Bessel region we have
\begin{equation}
\varphi\propto\sqrt{-u}J\Big(\frac{\sqrt{c}}{k},-\frac{p u}{k}\Big) \;,
\end{equation}
with asymptotic limit
\begin{equation}
\varphi\propto\cos\Big(px+\frac{\pi}{4}\Big[1+\frac{2\sqrt{c}}{k}\Big]\Big) \qquad u\to-\infty \;.
\end{equation}
By matching the different forms of the solution we find
\begin{equation}\label{J3}
|J|=\frac{1}{2\pi}\sqrt{\frac{q}{p}}\frac{\sqrt{c}}{k}\Gamma^2\Big(\frac{\sqrt{c}}{k}\Big)\bigg(\frac{4k^2}{pq}\bigg)^\frac{\sqrt{c}}{k}e^{S_U-\frac{2\sqrt{c}}{k}\ln U} \;.
\end{equation}
Thus the imaginary part of the effective action scales as
\begin{equation}
\text{Im }\Gamma\sim(1-\gamma^2)^{2(1+\sqrt{c}/k)} \;,
\end{equation}
which we recognize from our semiclassical results in
\cite{Gies:2015hia}. Thus, in contrast to the fields considered in the
previous sections with $n$ and $d>3$, weak fields decaying with
$|x|^{-3}$ have the same scaling in both the semiclassical and the
deeply critical regime.

Although $S_U$ diverges as $U\to\infty$, the limit of \Eqref{J3} is finite. Consider for example a Lorentz type of field,
\begin{equation}
f=\frac{u}{\sqrt{1+u^2}} \;,
\end{equation}
for which $c=1$. Expanding in $U$ leads to
\begin{equation}
S_U-\frac{2}{k}\ln U=\frac{\ln 4}{k}+\mathcal{O}(U^{-2}) \;.
\end{equation}
It follows from this, together with \Eqref{J3} and \Eqref{Gammarv},
that
\begin{equation}
\text{Im }\Gamma=\frac{L^2T\sqrt{\pi}}{16(2\pi)^2}E^\frac{3}{2}(1-\gamma^2)^{2(1+\frac{1}{E})}\bigg[\frac{e}{4}\bigg]^\frac{4}{E} \;,
\end{equation}
which, again, equals the critical limit of the semiclassical result in \cite{Dunne:2006st}. In general, the exponent in \Eqref{J3} can be expressed as
\begin{equation}
S_U-\frac{2\sqrt{c}}{k}\ln U=\frac{1}{k}\int\limits_{-U}^U\ud u\;\sqrt{1-f^2}-\theta(|u|-1)\frac{\sqrt{c}}{|u|} \;,
\end{equation} 
which makes it clear that the limit $U\to\infty$ is finite.

\section{Fields decaying as $E\sim |x|^{-2}$}\label{Fields decaying as x2}

In this section we will consider fields decaying as
\begin{equation}
E\to E_0\frac{c}{|kx|^2} \;.
\end{equation}
As for weak fields with $d=3$, we also recover the semiclassical
scaling as shown in the following. For $u\gg1$ we have
\begin{equation}
f=1-\frac{c}{u} \;,
\end{equation}  
and the Klein-Gordon equation reduces to
\begin{equation}
\Big(\partial_u^2+\frac{1}{k^2}\Big[-\frac{2c}{u}+q^2\Big]\Big)\varphi=0 \;.
\end{equation}
Instead of a Bessel function, this time the solution is given by a
Whittaker function\footnote{Whittaker functions also appear in
  treatments of pair production by constant electric fields in
  de~Sitter space, see
  e.g. \cite{Garriga:1993fh,Frob:2014zka,Kobayashi:2014zza,Stahl:2015gaa}
  and references therein.}
\begin{equation}\label{Whittaker1}
\varphi=\Big(\!-\frac{2iq}{k}\Big)^\frac{ic}{kq}W_{-\frac{ic}{kq},\frac{1}{2}}\Big[-\frac{2iqu}{k}\Big] \;.
\end{equation} 
The asymptotic limit $u\to\infty$ of \Eqref{Whittaker1} is not simply $e^{iqx}$, but
\begin{equation}\label{asymptotic-log}
\varphi\to\exp i\Big(qx-\frac{c}{kq}\ln (qx)+\text{real constant}\Big) \;.
\end{equation} 
However, we still have the same normalization
\begin{equation}
\varphi^\dagger(-i\overset{\leftrightarrow}{\partial})\varphi\to 2q \;,
\end{equation}
so we expect that the relation between $\mathcal{T}$ and $\text{Im }\Gamma$ 
derived in the literature is still valid.

\subsection{Weak fields}

We begin with the scaling for weak fields $E_0\ll1$. In the overlap
with the WKB regime $u\sim u_L^\LCp$, where
\begin{equation}
1\ll u\ll \frac{c}{k^2}\ll \frac{c}{q^2} \;,
\end{equation}
we use the asymptotic expansions in \cite{Buchholz:1969} for $W_{\kappa,\frac{1}{2}}(z)$
as $|\kappa|\to\infty$, yielding
\begin{equation}
\varphi=\sqrt{q}\Big(\frac{u}{2c}\Big)^\frac{1}{4}\exp\Big(\frac{\pi c}{kq}-\frac{2\sqrt{2cu}}{k}+i...\Big) \;.
\end{equation} 
By matching this with the WKB form \eqref{WKB3}, we find
\begin{equation}
|A|=\sqrt{q}\exp\Big(\frac{\pi c}{kq}-\frac{2\sqrt{2cU}}{k}\Big) \;.
\end{equation}
Already at this point, we can see the emergence of essential scaling
by recalling $q\sim\sqrt{1-\gamma^2}$. To the left of the WKB region
the solution is given by another Whittaker function
\begin{equation}
\varphi=b_2M_{\pm\frac{ic}{kp},\frac{1}{2}}\Big[\pm\frac{2ip(-u)}{k}\Big] \;.
\end{equation}
The asymptotic limit $u\to-\infty$ has the form of
  \Eqref{asymptotic} modified with logarithmic terms as in
\Eqref{asymptotic-log}. Using again expansions given in
\cite{Buchholz:1969}, we find
\begin{equation}\label{J2-weak}
|J|=\sqrt{\frac{q}{p}}\exp\Bigg[\frac{\pi c}{k}\Big(\frac{1}{p}+\frac{1}{q}\Big)+\left(\int_{-U}^U\!\frac{\sqrt{1-f^2}}{k}\right)-\frac{4\sqrt{2cU}}{k}\Bigg] .
\end{equation}
Substituting this into \Eqref{Gammarv} and taking $U\to\infty$, we
finally obtain 
\begin{equation}\label{Gammap2}
\begin{split}
\text{Im }\Gamma&=\frac{L^2T}{32\sqrt{6}\pi^3}\Big(\frac{k}{c}\Big)^\frac{3}{2}(1-\gamma^2)^\frac{11}{4}
\exp\left(-\frac{4\pi c}{k\sqrt{1-\gamma^2}}\right) \\ 
&\times\exp\left\{-\frac{2}{k}\!\left[\!\left(\int_{-U}^U\sqrt{1-f^2}\right)-{4\sqrt{2cU}}\right]\!\right\}\!\Bigg|_{U\to\infty}\!\!\! , 
\end{split}
\end{equation}
where the second line can asymptotically be expressed as 
\begin{equation}
\exp\left[-\frac{2}{k}\int\limits_{-\infty}^\infty\ud u\;\left(\sqrt{1-f^2}-\sqrt{\frac{2c}{|u|}}\right)\right] \;.
\end{equation}
This is indeed exactly the same scaling as we found in the
semiclassical regime \cite{Gies:2015hia}, which suggests that our
semiclassical results in \cite{Gies:2015hia} hold more generally in
the deeply critical regime for weak fields with $2\leq d\leq3$. For
the field implicitly defined by $f'=(1-f^2)^2$ the term in square
brackets in the second line in \Eqref{Gammap2} vanishes and one can
show that the prefactor in the first line of \Eqref{Gammap2} also
agrees with the semiclassical result.

\subsection{Strong fields}

\begin{figure}
\includegraphics[width=0.45\textwidth]{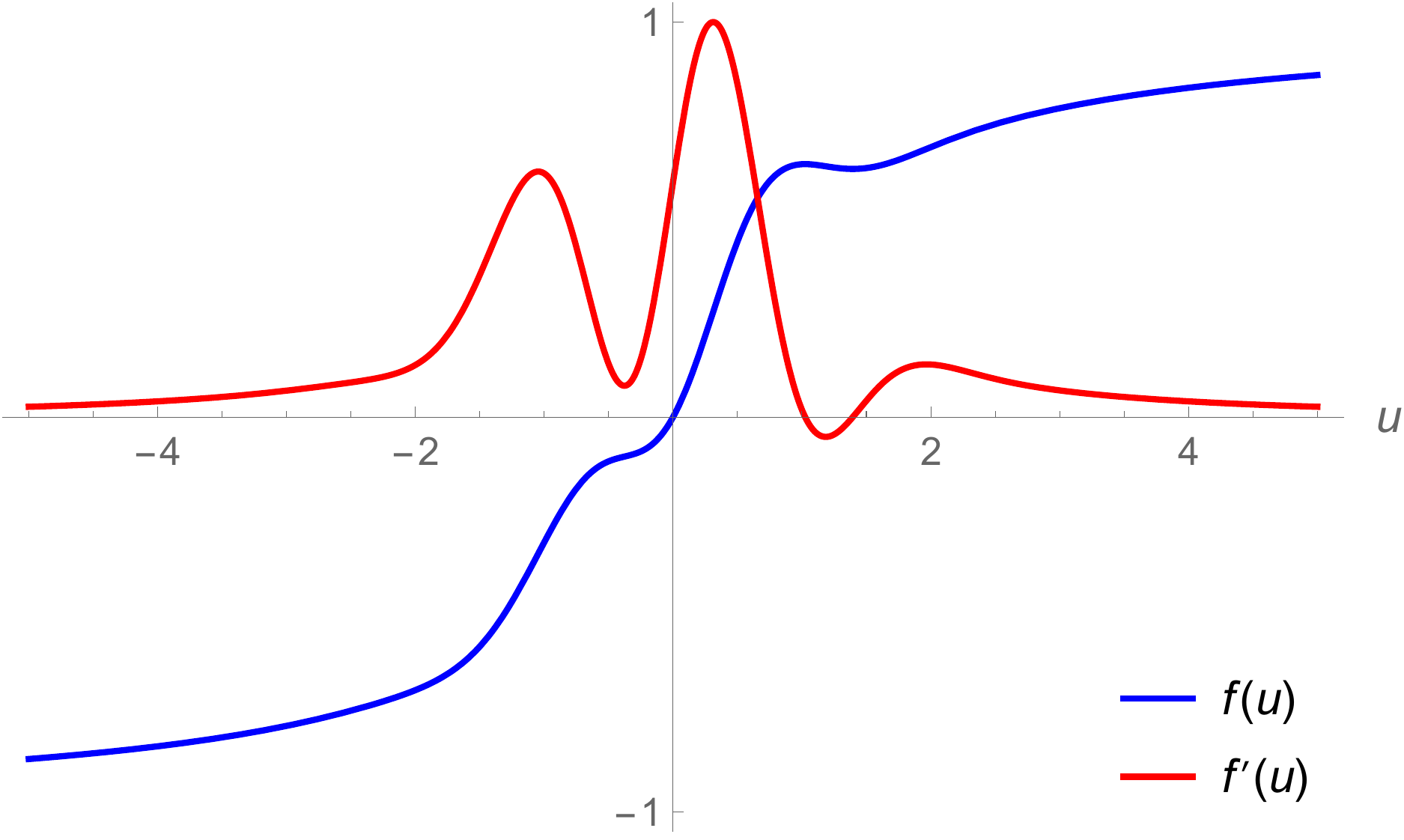}
\caption{Example field defined by~\eqref{fx2ex} that
  vanishes as $E\sim|x|^{-2}$; the blue curve is the potential $f(u)$
  and the red curve the normalized electric field $E/E_0=f'$.}
\label{nontrivial-x2-fig}
\end{figure}

\begin{figure}
\includegraphics[width=0.45\textwidth]{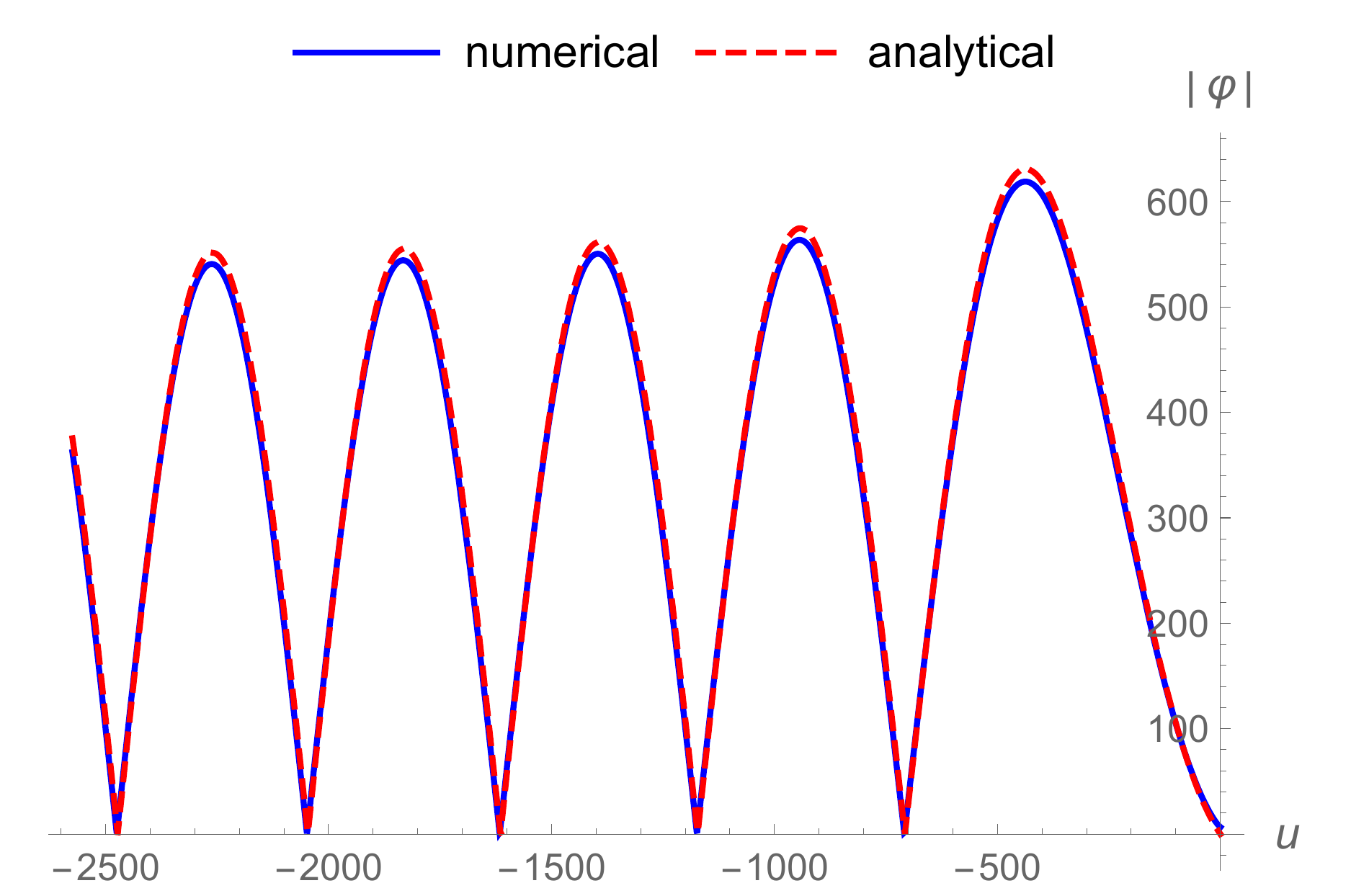}
\caption{Modulus of the solution to the Klein-Gordon equation
  $|\varphi(u)|$ in the left asymptotic region, i.e., for large
  negative $u$. The field shape is that shown in
  Fig.~\ref{nontrivial-x2-fig} with field strength $E_0=10$,
  adiabaticity $1-\gamma=3*10^{-3}$, and momentum parameters $r=0$ and
  $v=1$. The two curves show that the analytical results \eqref{LL2},
  \eqref{Whittaker-A} and \eqref{Whittaker-A-U-limit} agree well with
  the exact numerical solution; the analytical solution converges to
  the exact numerical one as $1-\gamma$ decreases.}
\label{abs_phi-nontrivial-x2-fig}
\end{figure}

Now let us study strong fields $E_0\gg1$. We consider again for
simplicity symmetric fields. Let $U$ be such that for $u>U$ we have
$f=1-c/u$. Although $U\gg1$ we still have $qU/k\ll1$, so in the region
$u\sim U$ we can expand \Eqref{Whittaker1}
\begin{equation}\label{LQ2}
\varphi=Q_0\Big(1+\frac{2cu}{k^2}\Big[\ln\Big(\frac{2cu}{k^2}e^
{2\gamma_E}\Big)-1\Big]\Big) \;,
\end{equation}
where $\gamma_E\approx0.577$ is the Euler number and
\begin{equation}
|Q_0|=\sqrt{\frac{kq}{2\pi c}}e^\frac{\pi c}{kq} \;.
\end{equation}
In the region $-U<u<U$ we solve the Klein-Gordon equation \eqref{Klein3} perturbatively in $1/k$,
\begin{equation}\label{LL2}
\varphi=Q_0\Big(1+\frac{2c}{k^2}\Big[c_0+c_1u+\int\limits_U^u\ud u'(u-u')\frac{1-f^2(u')}{2c}\Big]\Big) \;,
\end{equation} 
where the two constants $c_0$ and $c_1$ are obtained by matching
\Eqref{LL2} with \Eqref{LQ2} in the region $u\sim U$. For $u<-U$ the
solution is given by a linear combination of Whittaker functions,
\begin{equation}\label{MW}
\varphi=AM_{\frac{ic}{kp},\frac{1}{2}}\Big[-\frac{2ipu}{k}\Big]+BW_{\frac{ic}{kp},\frac{1}{2}}\Big[-\frac{2ipu}{k}\Big] \;.
\end{equation}
By matching \Eqref{MW} with \Eqref{LL2}, we find $B=\Gamma(1-ic/kp)Q_0$ and
\begin{equation}\label{Whittaker-A}
A=-\frac{icQ_0}{kp}\left[\left(\int_{-U}^U\frac{1-f^2}{2c}\right)-2\ln\frac{2cU e^{2\gamma_E}}{k^2}\right] \;,
\end{equation}
where the limit $U\to\infty$ of the term in square brackets is
finite and given by
\begin{equation}\label{Whittaker-A-U-limit}
\int\limits_{-\infty}^\infty\!\ud u\; \left(\frac{1-f^2}{2c}-\frac{\theta(|u|-1)}{|u|}\right)-2\ln\frac{2c e^{2\gamma_E}}{k^2} \;.
\end{equation}
In the asymptotic limit $u\to-\infty$ the first term in \Eqref{MW} is
dominant and the amplitude becomes
\begin{eqnarray}
|J|&=&\frac{1}{2\pi}\sqrt{\frac{q}{p}}\bigg|\left(\int_{-U}^U\frac{1-f^2}{2c}\right)-2\ln\frac{2cU e^{2\gamma_E}}{k^2}\bigg|
\nonumber\\
&&\times \exp\frac{\pi c}{k}\Big(\frac{1}{p}+\frac{1}{q}\Big) \;.\label{J2-strong}
\end{eqnarray}
To check these analytical expressions, we consider the field depicted in
Fig.~\ref{nontrivial-x2-fig}, and show in
Fig.~\ref{abs_phi-nontrivial-x2-fig} that the analytical approximation
for $|\varphi(u)|$ agrees well with the exact numerical solution.

By comparing \Eqref{J2-strong} with \Eqref{J2-weak} we see that the
critical scaling for strong fields is the same as that for weak
fields. The prefactor, though, is different. In fact, for very strong
fields we can drop the integral term in \Eqref{J2-strong},
\begin{equation}
|J|\approx\frac{2}{\pi}\sqrt{\frac{q}{p}}\ln k\exp\frac{\pi c}{k}\Big(\frac{1}{p}+\frac{1}{q}\Big) \;.
\end{equation}
Thus, for very strong fields vanishing as $E\sim|x|^{-2}$, the whole
expression for $\text{Im }\Gamma$ is universal and not just the
scaling behavior (c.f.~\Eqref{GammaSpinStrong}).

\section{Spinor QED}\label{Spinor QED}
 
So far we have considered scalar QED. In this section, we briefly
consider spinor QED \cite{Nikishov:1970br}. The imaginary part of the
effective action is still given by \Eqref{Gamma p-int}, but the tunneling factor now reads
\begin{equation}
\mathcal{T}=2\frac{p_0+p}{2/\gamma-p_0-q}\frac{q}{p}\frac{1}{|J|^2} \;,
\end{equation}
where the factor of $2$ comes from the sum over spin degrees of
  freedom. In the critical regime, this reduces to
\begin{equation}
\mathcal{T}=2\frac{q}{p}\frac{1}{|J|^2} \;,
\end{equation} 
which has the same form as the scalar case in \Eqref{scalar T}. The
amplitude $J$ is obtained from $\varphi$ as in \Eqref{asymptotic}, but
$\varphi$ is now the solution of the squared Dirac equation,
\begin{equation}\label{Dirac}
\Big(\partial_x^2+[p_0-A(x)]^2-m_\LCperp^2+iA'(x)\Big)\varphi=0 \;,
\end{equation}
which in the critical regime reduces to
\begin{equation}\label{Dirac2}
\left(\partial_u^2+\frac{1}{k^2}\Big[-(1-f^2)+ikf'+P^2]\right)\varphi=0 \;,
\end{equation}
where $P^2=q^2$ for $u\gtrsim0$, and $P^2=p^2$ for
$u\lesssim0$. Comparing with the scalar case we see that the $f'$ term
is new. It arises from the Pauli term $\sim \sigma_{\mu\nu}
  F^{\mu\nu}$ of the squared Dirac operator. For weak fields the
solution in the WKB region is
\begin{equation}\label{WKB spinor}
\varphi=\frac{A}{(1-f^2)^\frac{1}{4}}\exp\bigg(\frac{i}{2}\sin^{-1}f+\int\limits_u^\infty\frac{\sqrt{1-f^2}}{k}\bigg) \;.
\end{equation} 
For asymptotically decaying weak fields the new $f'$ term is always
smaller than the $1-f^2$ term: in the Bessel region we have
\begin{equation}
kf'\sim \frac{k}{u^d} \ll \frac{1}{u^d} \ll \frac{1}{u^{d-1}}\sim 1-f^2 \;,
\end{equation}
and in the overlap with the WKB region the extra term in \Eqref{WKB
  spinor} simply gives an irrelevant phase. Thus, scalar and spinor
QED agree in the deeply critical regime (up to a factor of $2$) for
asymptotically decaying weak fields. This is also reflected by the exact result for a Sauter pulse~\cite{Nikishov:1970br}, which for spinor QED leads to \Eqref{Gamma exp} times a factor of~$2$. 
For compact fields we have in the Bessel region
\begin{equation}
\frac{kf'}{1-f^2}\sim \frac{k}{u_\LCp-u} \;,
\end{equation}
so the spin term actually becomes larger than the scalar term as $u\to u_\LCp$, which suggests that the spin term can lead to corrections to the non-universal coefficient $D$. However, the spin term is only dominant in a region of size $u_\LCp-u\lesssim k$, so these corrections will not change $D$ significantly. Indeed, the most important part of the $D$ is still given by $e^S$ as in the scalar case. 
Moreover, the critical exponent remains unchanged,
\begin{equation}
\text{Im }\Gamma_\text{spin}\sim(1-\gamma^2)^3 \;.
\end{equation} 

\begin{figure}
\includegraphics[width=0.45\textwidth]{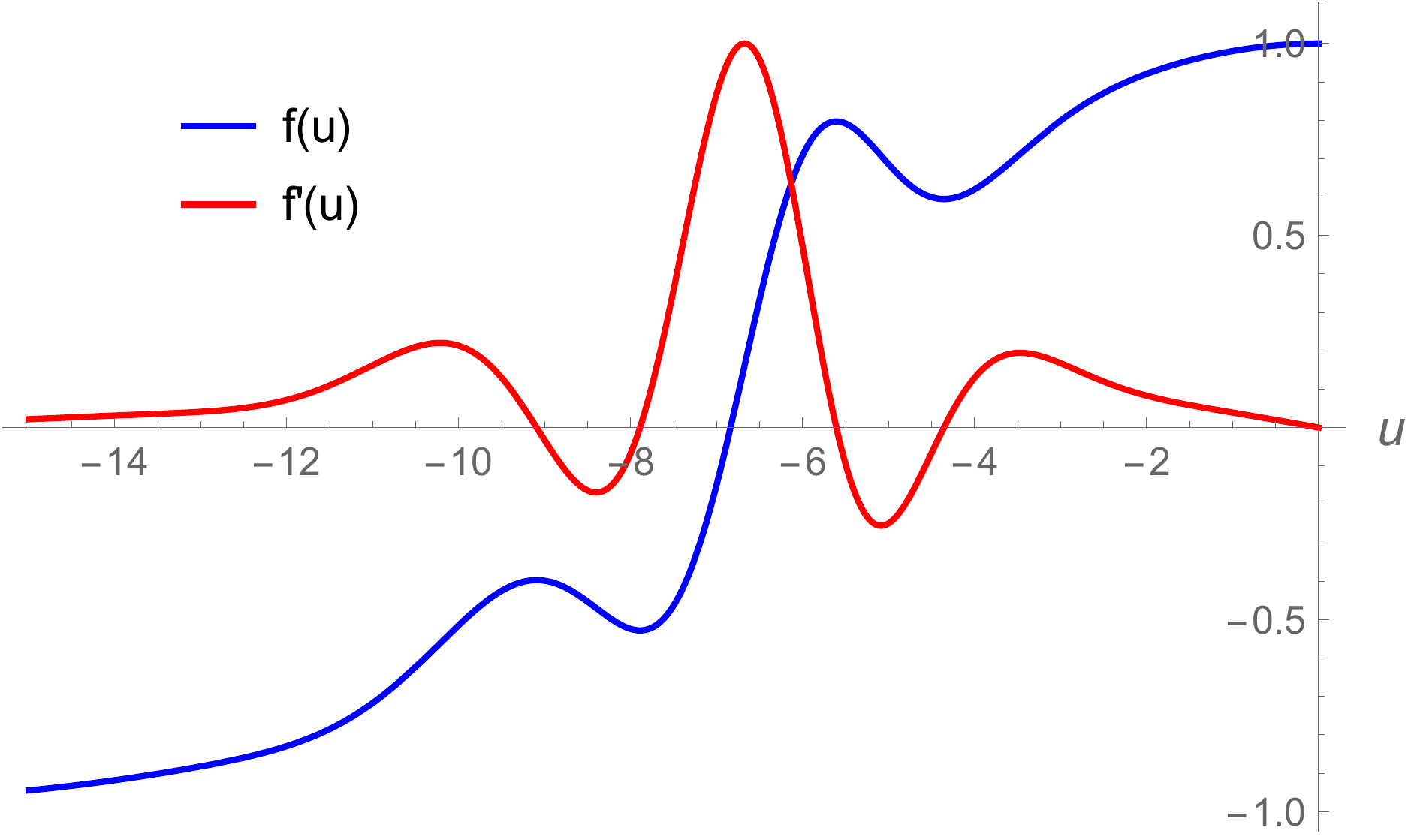}
\caption{Example field as defined in \Eqref{FieldExample12-def}. The blue curve depicts the potential
  function $f(kx)$ and the red curve is the electric field
  $E(kx)/E_0=f'$.}
\label{FieldExample12-fig}
\end{figure}

\begin{figure}
\includegraphics[width=0.45\textwidth]{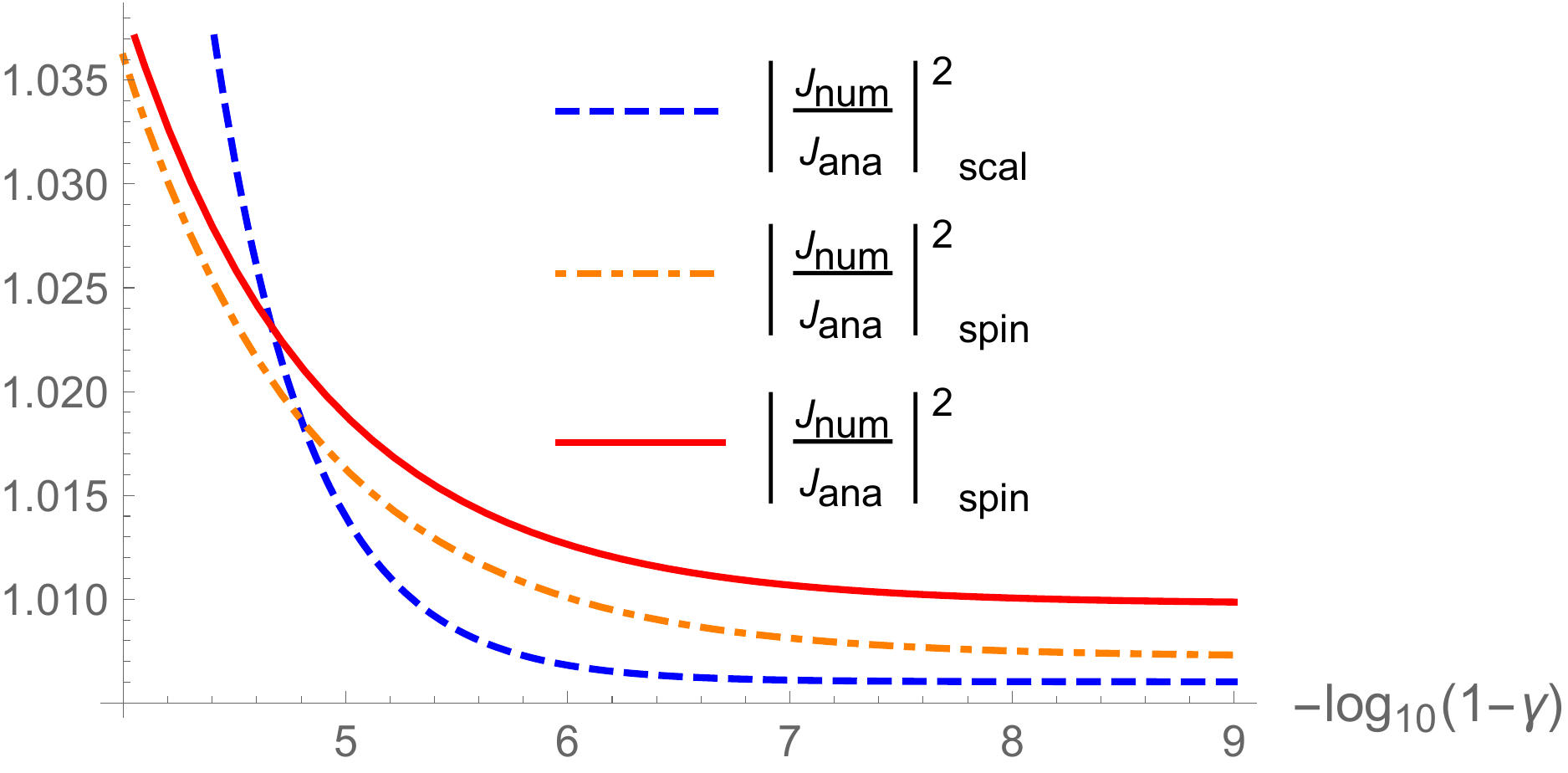}
\caption{Ratios of the squared amplitude obtained numerically and
  analytically for scalar and spinor QED. The field is shown
  in~\ref{FieldExample12-fig} with peak-field parameter $E_0=70$, and
  the momentum parameters are $r=0$ and $v=1$. For scalar QED,
  $J_\text{num}$ is obtained from the numerical solution
  of~\Eqref{Klein}, $J_\text{ana}$ is the analytical estimate of
  \Eqref{D-strong}, and their ratio approaches $\approx1-6*10^{-3}$.
  For spinor QED, $J_\text{num}$ is obtained from the numerical
  solution of~\Eqref{Dirac}. The latter is compared to two different
  analytical estimates $J_\text{ana}$ deduced
  from~\Eqref{D-strong-spin} and~\Eqref{JfromD} by either keeping
  (orange dot-dashed curve) or neglecting (red solid
  curve) the $1-f^2$ term, and their ratios approach
  $\approx1-7*10^{-3}$ and $\approx1-9.9*10^{-3}$, respectively.}
\label{SpinStrongNumAna-fig}
\end{figure}

\begin{figure}
\includegraphics[width=0.45\textwidth]{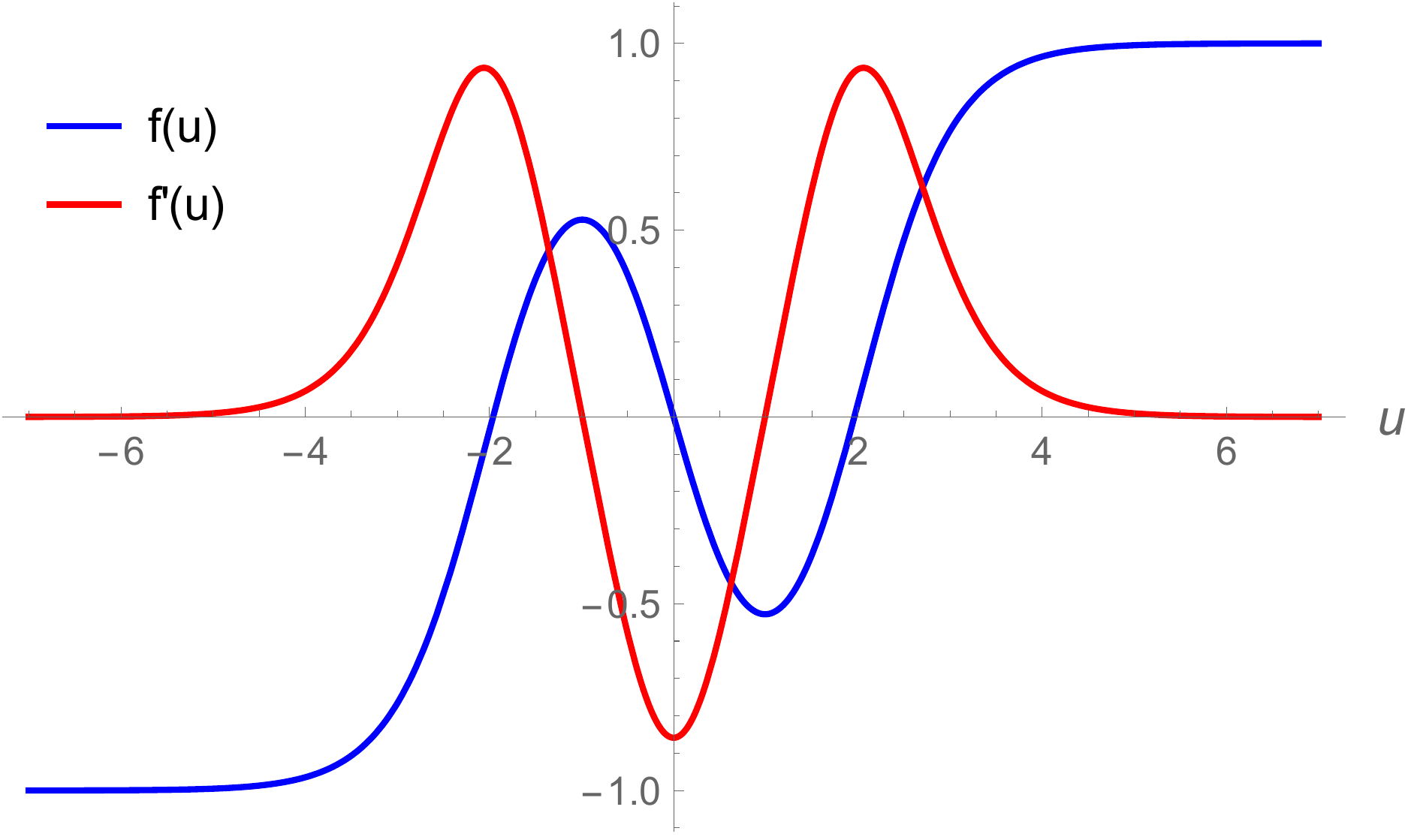}
\caption{Example field as defined in \Eqref{FieldExampleAsymptoticExpSym-def}. The blue curve depicts the potential
  function $f(kx)$ and the red curve is the electric field
  $E(kx)/E_0=f'$.}
\label{FieldExampleAsymptoticExpSym-fig}
\end{figure}

\begin{figure}
\includegraphics[width=0.45\textwidth]{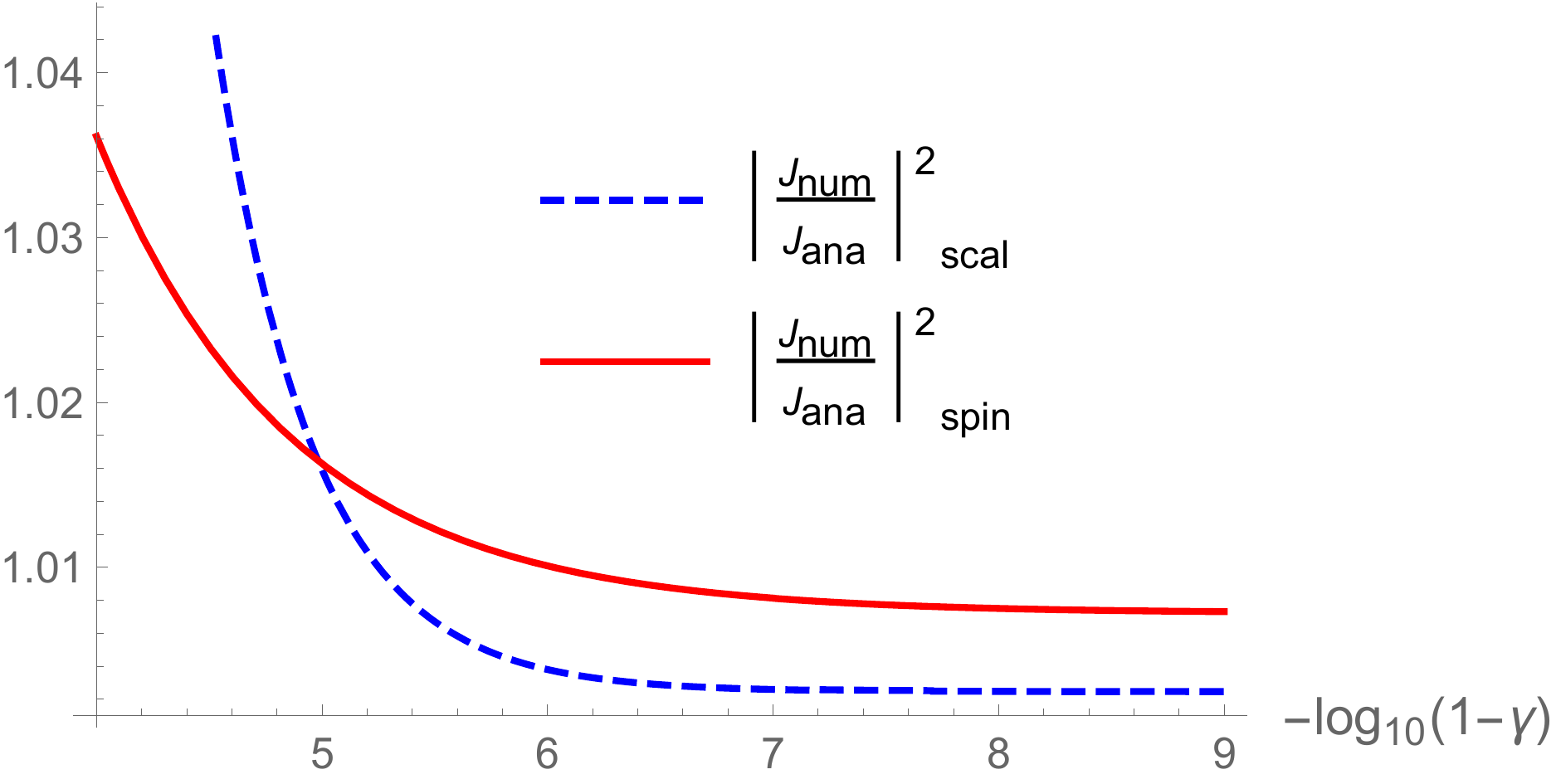}
\caption{Ratios of the squared amplitude obtained numerically and analytically for scalar and spinor QED, as described in Fig.~\ref{SpinStrongNumAna-fig}. The field is shown in~\ref{FieldExampleAsymptoticExpSym-fig}, the field strength is $E_0=70$ and the momentum parameters are $r=0$ and $v=1$. The scalar and spinor ratios converge to $\approx1-2.5*10^{-3}$ and $\approx1-7.3*10^{-3}$.}
\label{AsymptoticExpSymScalSpin-fig}
\end{figure}

For strong fields, a remarkable simplification occurs in the
  spinor case marking a qualitative difference to the scalar case: we
  can find $D_\text{spin}$ by going back to \Eqref{D-strong} and
  replacing $1-f^2$ with $1-f^2-ikf'$. While $1-f^2$ is bounded,
  the term $\sim k f'\sim E$ is proportional to the electric field,
  thus dominating the integrand in the strong-field limit,
\begin{equation}\label{D-strong-spin}
\begin{split}
D_\text{spin}&=\int\limits_{-\infty}^\infty\ud u\; \frac{1-f^2(u)-ikf'(u)}{k^2} \\
&\approx-\frac{i}{k}\int f'=-\frac{2i}{k}\;,
\end{split}
\end{equation}
and hence we find \Eqref{GammaSpinStrong}.
Equation \eqref{GammaSpinStrong} agrees with the exact result for a Sauter pulse~\cite{Nikishov:1970br} in the strong-field limit. In Figs.~\ref{SpinStrongNumAna-fig} and~\ref{AsymptoticExpSymScalSpin-fig}, we verify that~\Eqref{D-strong-spin} matches with the numerical solution of~\Eqref{Dirac} near the critical point for the fields shown in Figs.~\ref{FieldExample12-fig} and~\ref{FieldExampleAsymptoticExpSym-fig}, respectively.

The origin of this enhanced strong-field universality lies in the
dominance of the Pauli term $\sim\sigma_{\mu\nu}F^{\mu\nu}$,
parametrizing the coupling of the spin structure to the field. It is
instructive to recall the relevance of this term in the analog case of
a strong magnetic field: in the magnetic case, the Pauli term
characterizes the paramagnetism of the electron-positron fluctuations
as opposed to the Klein-Gordon operator describing diamagnetism. In
the strong-field limit, paramagnetism dominates which is reflected by
a zero-mode of the squared Dirac operator for the lowest Landau level
and a suitably oriented spin
\cite{Weisskopf:1996bu,Dittrich:1985yb}. This mode is responsible for
a variety of strong-field features that are unique to spinor QED
\cite{Tsai:1975iz,Gusynin:1994va,Gusynin:1994xp,Scherer:2012nn,Dobrich:2012sw,Dobrich:2012jd}. In
the present case, it is again the Pauli term dominating the strong
field limit, even though the Minkowskian mode structure in the
electric field is somewhat different from the magnetic case: e.g., there
  is no dependence on the orientation of the spin relative to the
  field, as the electron does not have a permanent electric
  dipole. Still, the analogy to the magnetic case justifies to
classify the enhanced universality of deeply critical pair production
in the spinor case as \textit{paraelectric dominance}. This
  nomenclature reflects the dominance of the endomorphism
  $\sim\sigma_{\mu\nu}F^{\mu\nu}$ in the relevant differential operator
  as opposed to the ``dia''-part of the covariant Laplacian~\cite{Nink:2012vd}.

\section{Conclusion}\label{Conclusion}

We have studied critical scaling in Schwinger pair production. In the
space of all possible electromagnetic field configurations, we have
considered the regime near the critical hypersurface that separates
the configurations that allow for pair production from those that
do not. This critical surface can be detected using the imaginary part
of the Heisenberg-Euler effective action $\text{Im }\Gamma$ as an
order parameter. As noted in our previous work \cite{Gies:2015hia},
the scaling of $\text{Im }\Gamma$ in the vicinity of the critical
surface exhibits scaling laws that are reminiscent to critical
phenomena in statistical systems. In the present paper, we generalize
these previous results in two decisive aspects: we provide for the
first time scaling results in the deeply critical regime, i.e. in the
immediate vicinity of the critical surface, whereas the results of
\cite{Gies:2015hia} apply to the semiclassical critical
region. Second, by directly studying the underlying field equations,
we can address a wider class of field profiles, also including
non-symmetric cases with non-monotonic gauge potentials. For
simplicity, we still restrict ourselves to uniaxial time-independent
field profiles varying in one spatial direction. 

In comparison with the semiclassical critical regime, the deeply
critical regime supports an even higher degree of universality, with
the same scaling law as a function of the Keldysh parameter $\gamma$,
\begin{equation} 
\text{Im }\Gamma\sim(1-\gamma^2)^\beta, \quad \beta=3, \label{eq:scaling}
\end{equation}
for all fields that asymptotically vanish faster than $|x|^{-3}$. The
result holds for both scalar and spinor QED in the weak- as well as
strong-field regime. The existence of such a scaling law expresses the
fact that the onset of pair production is dominated by the long-range
fluctuations of the electron-positron field and becomes insensitive to
the microscopic details of the field profile.

The highest degree of universality is found for spinor QED in the
strong-field regime, where paraelectric dominance also establishes a
universal prefactor. The scaling law \Eqref{eq:scaling} is modified
for more gradually vanishing fields. E.g., fields that vanish as
$|x|^{-2}$ obey the same scaling law as in the semiclassical
regime. The scaling for fields vanishing as $|x|^{-3}$ depends on the
field strength; for weak fields, we recover the semiclassical scaling,
whereas \Eqref{eq:scaling} holds again for strong fields.

As we noted in \cite{Gies:2015hia}, it is easy to generalize the
results for the spatially inhomogeneous fields considered here to
fields that depend on a linear combination of $x$ and $t$ as in
\cite{Ilderton:2015qda}. It would be interesting to investigate how
more nontrivial combinations of spatial and temporal inhomogeneities
affect the critical scaling found in this paper. Of particular
relevance for discovery experiments may be fields that support
dynamically assisted pair production \cite{Schutzhold:2008pz}, e.g.,
by adding a weak time-dependent field to the fields considered here,
e.g. as in \cite{Schneider:2014mla}.

Since the real and imaginary parts of correlation functions and thus
of the effective action are related by dispersion relations, the onset
of pair production in the critical regime will also leave an imprint
in the real part of the effective action. One physical manifestation
of such dispersion relations is, for instance, given by an anomalous
dispersion of a photon propagating through the field, see
e.g.~\cite{Heinzl:2006pn}, which can be extracted from the photonic
two-point correlator
\cite{Tsai:1975iz,Tsai:1975tw,Dittrich:2000zu,Karbstein:2013ufa,Meuren:2014uia}. However,
since the real part also receives contributions from perturbative
processes, it remains an open question as to whether the analogue of
the (nonperturbative) scaling behavior can be quantitatively dominant
in the regime of anomalous dispersion.

Dispersion relations also lead to an intriguing relation between the
imaginary part of the action and the large-order behavior of the
perturbative expansion of the real part: for constant fields as well
as for the spatially homogeneous electric Sauter profile in time, it
has been shown that the absence/appearance of the imaginary part is
tightly linked to the properties of the perturbative series under
Borel transformations
\cite{Dunne:1999uy,Dunne:2002qg,Dunne:2004nc,Dunne:2004xk} (see
also~\cite{Heinzl:2006pn}). If this pattern also holds for spatially
inhomogeneous fields, the critical surface of Schwinger pair
production in the space of field profiles could also separate Borel-
from non-Borel-summable perturbative expansions of the effective
action. Such a conjecture clearly deserves further investigation.

\acknowledgments

We thank Gerald Dunne, Anton Ilderton, Felix Karbstein, Ralf
Sch\"utzhold, Johannes Oertel and Christian Schneider for interesting discussions.  G.T. thanks TPI, FSU Jena,
and HI Jena for hospitality during a research visit.  We acknowledge
support by the BMBF under grant No.  05P15SJFAA (FAIR-APPA-SPARC)
(H.G.), and the Swedish Research Council, contract 2011-4221 (P.I.: A. Ilderton) (G.T.).

\section*{Appendix}

Here we list the field examples used above to demonstrate various results. The first field example, shown in Fig.~\ref{ex1}, is given by $f(u)=f_0(u/\hat{f}_0')$, where $f(u>0)=1$,
\be\label{fexample1}
\begin{split}
f_0(u<0)=&1+\Big(1+\frac{1}{5}\sin(13u)\text{sech}^2\Big[3u+\frac{21}{10}\Big]\Big) \\
&\frac{4}{3\pi}\Big[\frac{u(-3+8u^2+3u^4)}{(1+u^2)^3}+3\arctan u\Big]
\end{split}
\ee
and where $\hat{f}_0'\approx4.5$ is the maximum of $f_0'$. 
The second field example, shown in Fig.~\ref{ex2}, is given by
\be\label{fexample2}
f(u)=\tan\Big(\frac{u}{3}\Big)-\frac{1}{5}\cos\Big(\frac{10u}{3}\Big)\exp-\Big(\frac{u}{3}\Big)^2 \;.
\ee 
The compact example in Fig.~\ref{nontrivial-weak-compact-amplitude-fig} is obtained by first integrating
\be
f_0'(u)=-u\cos\Big(u+\frac{\pi}{2}\Big)\Big(1+\frac{1}{4}\cos\Big[5\Big(u+\frac{\pi}{2}\Big)\Big]\Big)
\ee
and then
\be\label{fCompactEx}
f(-\pi<u<0)=1-2\frac{f_0(0)-f_0(u)}{f_0(0)-f_0(-\pi)} \;,
\ee
$f(u<-\pi)=-1$ and $f(u>0)=1$. The field example Fig.~\ref{abs_phi-nontrivial-x2-fig} is given by $f(u)=f_0(u/\hat{f}_0')$, where
\be\label{fx2ex}
f_0(u)=\frac{2}{\pi}\arctan\Big[\frac{\pi}{2}u\Big]\Big(1+\sin(3u)\exp(-3u^2)\Big)
\ee
and where $\hat{f}_0'\approx1.7$ is the maximum of $f_0'$.
An example similar to~\eqref{fexample1}, but with simpler analytical form, is given by $f(u)=f_0(u/\hat{f}_0')$, where $f_0(u>0)=1$,
\be\label{FieldExample12-def}
f_0(u<0)=1-\frac{2u^2}{(1+u^6)^\frac{1}{3}}\Big(1+\frac{7}{10}\sin[14u]\text{sech}^2[5u+3]\Big)
\ee
and where $\hat{f}_0'\approx10$ is the maximum of $f_0'$. \Eqref{FieldExample12-def} is shown in Fig.~\ref{FieldExample12-fig}.
The symmetric field shown in Fig.~\ref{FieldExampleAsymptoticExpSym-fig} is defined by
\be\label{FieldExampleAsymptoticExpSym-def}
f(u)=\tanh(u+2)-\tanh(u)+\tanh(u-2) \;.
\ee

\bibliography{bibliography}

\end{document}